\documentclass[11pt]{iopart}
\usepackage[unicode=true, bookmarks=false, breaklinks=false, pdfborder={0 0 1},colorlinks=false]{hyperref}
\usepackage{cite}
\usepackage{graphicx}
\usepackage{wrapfig}

\begin{document}

\title[From electronic structure to magnetism and skyrmions (Topical review)]{From electronic structure to magnetism and skyrmions (Topical review)}

\author{Vladislav Borisov}

\address{Department of Physics and Astronomy, Uppsala University, Box 516, SE-75120 Uppsala, Sweden}
\ead{vladislav.borisov@physics.uu.se}
\vspace{10pt}
\begin{indented}
\item[]October 2023
\end{indented}

\begin{abstract}
Solid state theory, density functional theory and its generalizations for correlated systems together with numerical simulations on supercomputers allow nowadays to model magnetic systems realistically and in detail and can be even used to predict new materials, paving the way for more rapid material development for applications in energy storage and conversion, information technologies, sensors, actuators etc. Modelling magnets on different length scales (between a few $\mathrm{\AA}$ngstr\"om and several micrometers) requires, however, approaches with very different mathematical formulations. Parameters defining the material in each formulation can be determined either by fitting experimental data or from theoretical calculations and there exists a well-established approach for obtaining model parameters for each length scale using the information from the smaller length scale. In this review, this approach will be explained step-by-step in textbook style with examples of successful multiscale modelling of different classes of magnetic materials from the research literature as well as based on results newly obtained for this review.
\end{abstract}

\vspace{2pc}

\newcommand{\unit}[2][1]{#1~\mathrm{#2}}

\section{Introduction}

Materials science in the second half of the twentieth century has been marked by the development of theoretical methods and application of supercomputers for simulating and studying properties of real materials. One of the major breakthroughs of theory was the density functional theory (DFT) suggested by Hohenberg and Kohn~\cite{Hohenberg1964} and the subsequent creation of the special \textit{ansatz} by Kohn and Sham~\cite{Kohn1965}. DFT allows to describe, to a good approximation, the complex correlated behavior of electrons which determine most of the material properties. The Kohn-Sham ansatz suggests a practical way of doing so when a sufficiently accurate form of electronic correlation energy is constructed. These developments created a flurry of activities in theoretical materials science aimed at finding better approximations for electronic correlations and more efficient numerical methods for solving the DFT equations. During the decades after the seminal Hohenberg-Kohn and Kohn-Sham papers, DFT has proved to be an efficient and reliable theory for describing realistic material models and there are many success stories where DFT calculations played a major role in understanding a certain physical phenomena or designing new functional materials (for examples see \cite{Sholl2009,Lebegue2013,Hasnip2014}).

At this level of theory, all the details of the crystal structure are fully taken into account and the relevant quantities obtained in calculations are the electronic wavefunction $\Psi(\{r_i\})$, density $n(r)$ and energy spectrum $\varepsilon_n(k)$ as a function of the wavevector $k$. The computational complexity of DFT calculations scales, at least, as $N^3$ where $N$ is the number of electrons/atoms in the unit cell and DFT simulations are usually restricted to length scales of up to several nanometers. Numerical methods optimized for calculating electronic structure in real space \cite{Peduto1997} have been developed to tackle the DFT equations for a few thousand \cite{RESCU2016} or even several hundred thousand atoms \cite{Dogan2023}, but this does not allow to go much beyond the aforementioned length scale.

Information about the nanometer-scale behavior is, however, not enough for understanding the magnetic phenomena in real materials, since new behavior can emerge on larger length scales. To circumvent the difficulty of DFT and quantum-mechanical methods, in general, to describe large groups of atoms, it is customary to construct \textit{effective models}, which are supposed to describe a subset of materials properties. In the context of magnetism, a spin model with Heisenberg $J_{ij}$ and Dzyaloshinskii-Moriya (DM) $\vec{D}_{ij}$ interactions and on-site anisotropy (for this example, uniaxial anisotropy of magnitude $K_U$) is the most famous example:
\begin{equation}
    H = -\sum\limits_{\left\langle ij \right\rangle} J_{ij} (\vec{S}_i \cdot \vec{S}_j) -\sum\limits_{\left\langle ij \right\rangle} D_{ij} (\vec{S}_i \times \vec{S}_j) + \sum\limits_i K_U (\vec{S}_i\cdot\vec{e})^2
    \label{e:Heisenberg_model}
\end{equation}

\begin{figure}
{\centering
\includegraphics[width=0.99\textwidth]{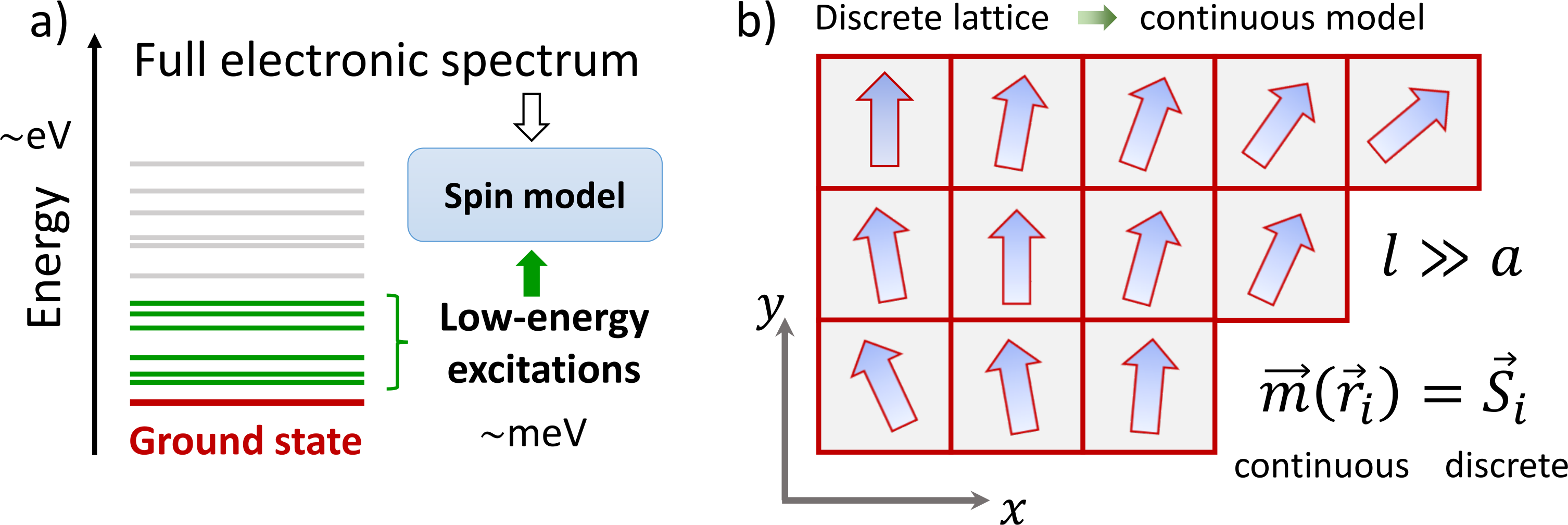}
}
\caption{a) Main idea of the effective spin model. Low-energy part of the full electronic spectrum is used to construct a model that describes excitations from the magnetic ground state. b) Illustration of the micromagnetic approach where the material is described by a continuous magnetization $\vec{m}(\vec{r})$, instead of discrete lattice of spins $\vec{S}_i$ sitting at positions $\vec{r}_i$. The system is subdivided into small regions $\sim\!\unit[1]{nm}$ with the local magnetization (shown by arrows) representing the average of atomic spins in that volume.}
\label{f:effective_models}
\end{figure}
Instead of describing the whole electronic energy spectrum and all the electronic degrees of freedom, this generalized Heisenberg model and models similar to it focus instead on the \textit{low-energy part of the spectrum}, close to the ground state (Fig.~\ref{f:effective_models}a). The electronic spin density is assumed to be well localized around each atom ($i$), so that its norm is constant and only the direction of spins (indicated by unit vectors $\vec{S}_i$) changes as a result of electron dynamics. In that case, expression (\ref{e:Heisenberg_model}) describes the energy of excited states where spins on different atoms deviate from the magnetic ground state, which includes spin waves and single-atom excitations. The advantage of the Heisenberg model is that parameters $J_{ij}$ contain a large part of complexity of electronic properties and the model itself is easier to solve than the equations of the density functional theory. This allows to perform simulations for up to $10^8$ spins, which covers the length scales up to several hundred nanometers relevant for many interesting magnetic states, such as spin spirals, domain walls, skyrmions, vortices etc. \cite{uppasd,Eriksson2017} (more details in Sec.~\ref{t:LLG_equations}).

Further increase of the length scale of accessible magnetic phenomena is possible by means of micromagnetic simulations where the discrete atomistic character of material is neglected and instead the continuous description of magnetic properties is applied. The main idea of the micromagnetic approach is illustrated in Fig.~\ref{f:effective_models}b, where a continuous function $\vec{m}(\vec{r})$ describes the local magnetization in different regions of the system and replaces the information about spin vectors $\vec{S}_i$ on different atomic sites at positions $\vec{r}_i$, while ideally $\vec{m}(\vec{r}_i)=\vec{S}_i$ should be satisfied. This approach works well when the magnetization varies on the length scale ($l$) much larger than the characteristic atomic distance ($a$), so that the nearest-neighbor spins have similar orientations. Magnetic ground state can be determined by minimizing the micromagnetic functional $E = \int \varepsilon(\vec{r})\,\mathrm{d}V$ with the energy density $\varepsilon \equiv \varepsilon(\vec{r})$:
\begin{equation}
    \varepsilon = A \left[ m_x \nabla^2 m_x + m_y \nabla^2 m_y + m_z \nabla^2 m_z \right] + \vec{m}\cdot ( \hat{D}\,\vec{\nabla} ) \times \vec{m} + K_U (\vec{m}\cdot\vec{e})^2
    \label{e:micromagnetic_functional}
\end{equation}

The two terms in this functional directly correspond to the Heisenberg and DM interactions in the spin model (\ref{e:Heisenberg_model}) and the parameters, spin stiffness $A$ and Dzyaloshinskii-Moriya matrix $\hat{D}$, can be determined from the interatomic $J_{ij}$ and $\vec{D}_{ij}$ interactions, while the uniaxial anisotropy has the same form but different units in the atomistic and micromagnetic cases. In the following (Sec.~\ref{t:micromagnetics}), we will discuss in detail how the micromagnetic functional (\ref{e:micromagnetic_functional}) is derived.

Both in atomistic and micromagnetic approaches described by models (\ref{e:Heisenberg_model}) and (\ref{e:micromagnetic_functional}), the magnetization dynamics can be studied by calculating the effective field $\vec{B}_\mathrm{eff}$ acting on each spin or micromagnetic element and integrating the Landau-Lifshitz-Gilbert equations (further details in Sec.~\ref{t:LLG_equations}). The micromagnetic method allows to approach the length scales of up to a few micrometers, which is necessary, for example, for studying wide domain structures and relatively large groups of topological magnetic objects, which would be hard to model using the atomistic spin dynamics. However, the atomistic spin dynamics is a more general and accurate way of modelling magnetic properties in and out of equilibrium and can be applied to any type of magnetic system (collinear and non-collinear magnets with multiple sublattices).

\begin{figure}
{\centering
\includegraphics[width=0.99\textwidth]{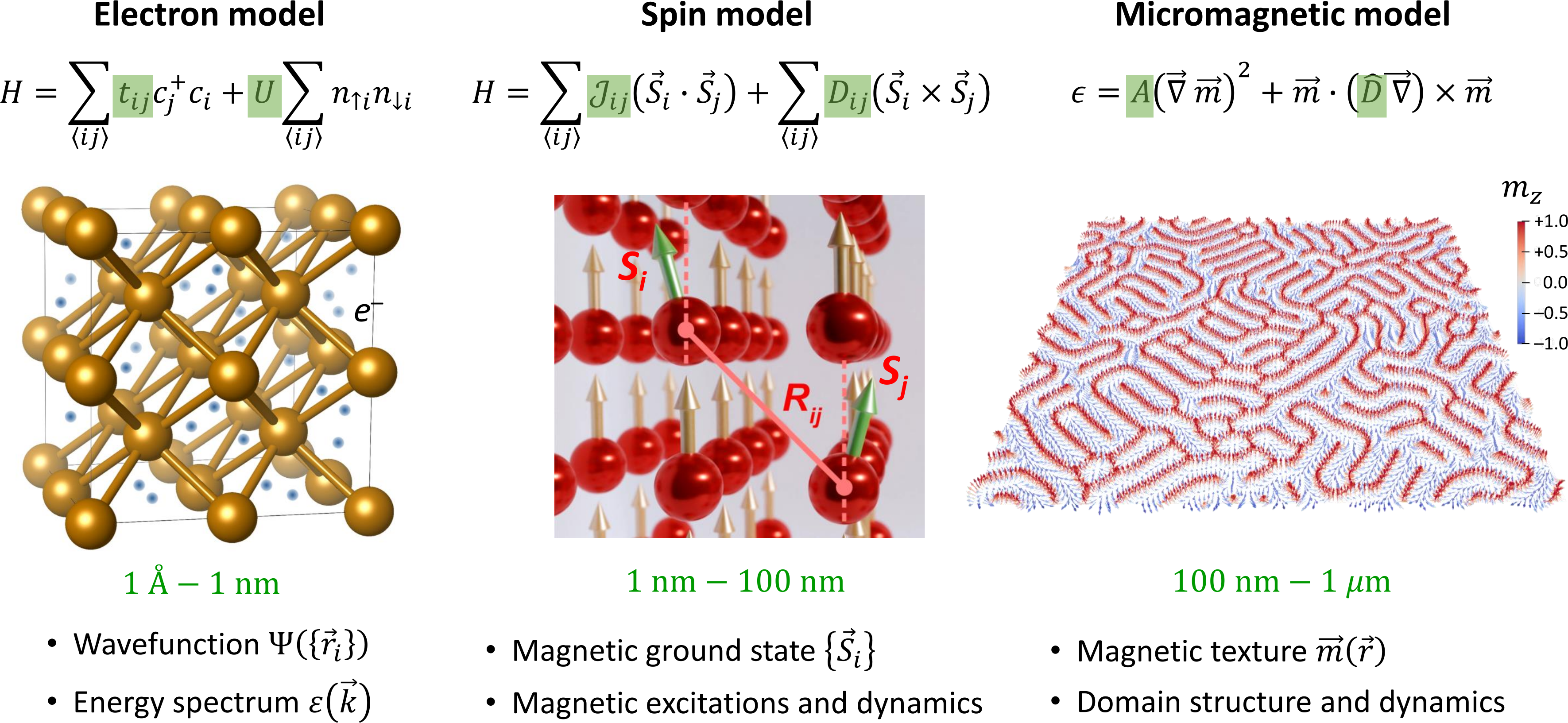}
}
\caption{Schematic illustration of the multiscale approach for modelling magnetic systems. First, the electronic properties are calculated from the first principles of quantum mechanics, leading to electron model (details in Sec.~\ref{e:elec_str}). From this, an effective spin model can be obtained, characterized by Heisenberg $J_{ij}$ and Dzyaloshinskii-Moriya $D_{ij}$ parameters that describe the interactions between spins $\vec{S}_i$ and $\vec{S}_j$ on different atomic sites (details in Sec.~\ref{t:spin_models}). Finally, micromagnetic description of magnetic texture (example of a typical simulation result is shown) is derived from the atomistic spin model of the previous step (detals in Sec.~\ref{t:micromagnetics}). The orders of magnitude of the relevant length scales and some of the material properties are listed for each model, together with the basic equations in the simplified form. Model parameters are highlighted by the green color.}
\label{f:multiscale_approach}
\end{figure}

From the introduction above, it is clear that there is a possibility of \textit{multiscale modelling} of magnetic systems following three basic steps illustrated in Fig.~\ref{f:multiscale_approach}:
\begin{itemize}
    \item \textbf{Electronic structure:} the electronic energy spectrum and related properties (hopping parameters $t_{ij}$, orbital character of the wavefunction) are calculated from first principles for a given crystal structure and chemical composition;
    \item \textbf{Atomistic spin model:} the energy of the system is parameterized using a spin model and the magnetic interactions are calculated based on the electronic properties;
    \item \textbf{Micromagnetic description:} discrete atomic structure of the system is neglected and a continuous magnetic model is derived from the atomistic model under the assumption of slowly varying magnetization.
\end{itemize}

In the rest of the review, different steps of this multiscale approach will be explained in general and a few representative examples from the literature will be discussed. Most of the examples are concerned with systems having a significant Dzyaloshinskii-Moriya interaction which stabilizes non-collinear or even topological textures. Importantly, this review does not aim at providing a comprehensive picture of different magnetic phenomena that can emerge in electronic system but rather focuses on the fundamental aspects of magnetism and general theoretical methodology for predictive modelling of any given magnetic system fully from first principles.

\section{Electronic structure}
\label{e:elec_str}

Soon after the foundations of quantum mechanics were laid in 1920's, it became clear that many properties of materials (conductivity, mechanical strength, magnetism etc.) are determined by the behavior of electrons in the crystal lattice. It was also clear that understanding this behavior is a formidable task for theory due to the long-range Coulomb interactions between all the electrons in the system, which decay rather slowly as $1/r$ ($r$~-- distance between electrons) and can lead to complex correlation effects. Even nowadays, the full picture of electronic correlations in solids is missing, although there has been a lot of progress in theoretical description of correlations using different approaches. One such approach is based on \textit{density functional theory} (DFT), which was formulated by Pierre Hohenberg and Walter Kohn in 1964 \cite{Hohenberg1964} and has become a widely used and very successful tool for modelling realistic materials. In this section, general overview of the DFT methodology is given with a focus on how it can be used to understand magnetic systems.

\subsection{Main idea of DFT}

In principle, the quantum-mechanical model of a solid crystal is well defined in a sense that the exact form of the Hamiltonian is known:
\begin{equation}
    \hat{H} = -\sum\limits_{i=1}^N \frac{\hbar^2 \nabla_i^2}{2m} +\sum\limits_{i<j} \frac{e^2}{|\vec{r}_j - \vec{r}_i|} -\sum\limits_{i,k} \frac{Z_k e^2}{|\vec{r}_i - \vec{R}_k|} -\sum\limits_{k=1}^{N_0} \frac{\hbar^2 \nabla^2_k}{2M_k} +\sum\limits_{k<k'} \frac{Z_k^2 e^2}{|\vec{R}_k - \vec{R}_{k'}|}
    \label{e:full_Hamiltonian}
\end{equation}

This Hamiltonian contains i) the kinetic energy of $N$ electrons each with a mass $m$ and charge $e$, ii) the Coulomb repulsion between them, iii) their electrostatic attraction to the nuclei with atomic numbers $Z_k$, iv) the kinetic energy of the $N_0$ nuclei with masses $M_k$, and v) their electrostatic repulsion energy. In many cases, the Born-Oppenheimer approximation \cite{Born1927} is valid, which assumes that the electron dynamics is much faster than the nuclei dynamics. One can write then a purely electronic Hamiltonian for given positions of the nuclei:
\begin{equation}
    \hat{H}(\{\vec{R}_k\}) = -\sum\limits_{i=1}^N \frac{\hbar^2 \nabla_i^2}{2m} +\sum\limits_{i<j} \frac{e^2}{|\vec{r}_j - \vec{r}_i|} -\sum\limits_{i,k} \frac{Z_k e^2}{|\vec{r}_i - \vec{R}_k|}
    \label{e:electronic_Hamiltonian}
\end{equation}

However, even this simplified quantum-mechanical problem is still virtually impossible to solve, since the number of electrons in a solid can approach astronomical numbers. Already a relatively small sample with a size of $\unit[(10\times10\times10)]{nm}$ can contain around $\sim\!\!10^{6}$ electrons. Even storing the information about the full electronic wavefunction would be beyond the capabilities of modern computers, considering that the amount of required memory scales exponentially with the number of electrons.

Alternative approach, \textit{density functional theory} (DFT), was suggested in the seminal paper by Hohenberg and Kohn \cite{Hohenberg1964}, who proposed to use the electronic density $n(\vec{r})$ instead of the full correlated wavefunction $\Psi$ to describe electronic systems. In the paper, it is proved that the energy of the system can be expressed as a functional of the density only and the ground state electron density minimizes this energy functional. This greatly simplifies the task of solving the electronic problem, because i) density is a function of 3 arguments (components of $\vec{r}$), while the original wavefunction depends on the positions $\{ \vec{r}_i \}$ of all electrons, ii) the ground state can be found by variational principle. The later aspect has been further elaborated on in the paper by Kohn and Sham \cite{Kohn1965} who introduced a special \textit{ansatz} to reformulate the DFT equations using an auxiliary independent-particle system. Each auxiliary particle has a wavefunction $\psi_i$ and experiences an effective potential which contains contributions from different interactions appearing in Eqn.~\ref{e:full_Hamiltonian} (ionic potential $V$, Hartree potential related to density $n(\vec{r})$ and the so-called exchange-correlation potential $V_{xc}$):
\begin{equation}
    \left[ -\frac{\hbar^2}{2m}\nabla^2 + V(\vec{r}) + \int \frac{n(\vec{r}\,')}{|\vec{r} - \vec{r}\,'|}\mathrm{d}\vec{r}\,' + V_{xc}(\vec{r}) \right] \psi_i(\vec{r}) = \epsilon_i \psi_i(\vec{r})
    \label{e:KS_equations}
\end{equation}

These independent particles have nothing to do with original electrons, neither relate their wavefunctions $\psi_i$ to the original wavefunction $\Psi$ of the full electron system described by Hamiltonian (\ref{e:full_Hamiltonian}). However, the actual electron density is postulated in the Kohn-Sham theory to correspond to the total density of auxiliary particles:
\begin{equation}
    n(\vec{r}) = \sum\limits_{i=1}^N |\psi_i(\vec{r})|^2
    \label{e:KS_density}
\end{equation}

This density also enters each of the $N$ Kohn-Sham equations (\ref{e:KS_equations}), meaning that these have to be solved self-consistently with Eqn.~\ref{e:KS_density}. The solution is usually found iteratively by updating the electronic density and DFT potential after each step until both quantities change negligibly based on the convergence criterion.

\subsection{DFT approximations}
In the density functional theory, all the correlations between electrons are encoded in the exchange-correlation potential $V_{xc}(\vec{r})$, the exact expression for which is unknown but has been a subject of intense research over many decades. In actual calculations for real materials one has to use approximations for this functional, of which there are several hundreds. Historically, the first one was the \textit{local-density approximation} (LDA) \cite{Kohn1965,Hedin1965} where the local correlation energy density $\varepsilon_{xc}(n)$ is postulated to coincide with that of a homogeneous electron gas of the same density. The later has been calculated using stochastic simulations of the quantum-mechanical electronic problem \cite{Ceperley1980}. Accordingly, the exchange-correlation energy functional $E[n]$ and potential $V_{xc}(\vec{r})$ in LDA read:
\begin{equation}
    E[n(\vec{r})] = \int \mathrm{d}^3 r\, \varepsilon_{xc}(n) \,n(\vec{r}), \hspace{10pt} V_{xc}(\vec{r}) = \frac{\delta E[n]}{\delta n} = \varepsilon_{xc}(n) + n\, \frac{\partial \varepsilon_{xc}(n)}{\partial n}  \label{e:LDA_functional}
\end{equation}

In this form, the local-density approximation does not contain any spin dependence and $n(\vec{r})$ is the total electron density. Importantly, the LDA approach can be extended to magnetic systems, which was done by Barth and Hedin \cite{Barth1972}. The DFT potential now becomes spin-dependent which allows solutions of DFT equations with non-equal spin-up and spin-down densities $n^\uparrow$ and $n^\downarrow$, such that the total density is $n(\vec{r}) = n^\uparrow(\vec{r}) + n^\downarrow(\vec{r})$ and the magnetization is $m(\vec{r}) = n^\uparrow(\vec{r}) - n^\downarrow(\vec{r})$:
\begin{equation}
    V_{xc}^\uparrow = F_1[n] \sqrt[3]{n^\uparrow(\vec{r})} + F_2[n], \hspace{5pt} V_{xc}^\downarrow = F_1[n] \sqrt[3]{n^\downarrow(\vec{r})} + F_2[n] \label{e:LSDA_potential}
\end{equation}

Despite assuming a slowly varying electron density, the LSDA approach has been quite useful for calculating the electronic properties of different types of solid-state systems, both in bulk and near surfaces and interfaces where the density is far from being uniform. Also, the LSDA predictions for the magnetic properties (magnetic moments and interactions, as we discuss in Sec.~\ref{t:spin_models} later on) are quite accurate in many cases, for example, for elemental transition metals Fe, Co and Ni and rare earths as well as for weakly correlated transition metal compounds (for more detailed discussion see \cite{Eriksson2017}). Moreover, LSDA correctly predicts that all other elements are non-magnetic in a sense that they do not have intrinsic moments, while Pd is close to being magnetic due to the so-called \textit{Stoner instability} \cite{Stoner1933}. This instability is triggered when the electronic density of states at the Fermi level $n(E_\mathrm{F})$ is large enough, so that it becomes favourable for electrons to attain a finite spin polarization which reduces the total energy (more discussion of fundamentals in \cite{Blundell2001,Kubler2021}). The Stoner model is useful for understanding the origin of magnetism in solids in simple terms. It contains the kinetic energy of electrons, which increases when a finite magnetization is induced due to increase of Fermi energy, and magnetic exchange energy which is negative and modelled as $E_m = -m\cdot I$, where $m$ is the total magnetization and $I$ is the Stoner parameter that can be calculated from density functional theory, which was done, for example, for 3\textit{d} and 4\textit{d} elements in \cite{Janak1977}. The competition between the two energies determines whether a system is magnetic or not, and the Stoner term $-m\cdot I$ is a simple model of complex electronic correlation effects which, together with quantum statistics for the wavefunction (more details in Sec.~\ref{t:spin_models}), can lead to the emergence of magnetism. The condition for this is $n(E_\mathrm{F})\cdot I > 1$ which marks the onset of the aforementioned Stoner instability where the gain in magnetic energy outweighs the kinetic energy growth.

A natural way to try to improve the local-density approximation is to take into account the density variations in the first order by including the density gradient in the DFT potential, leading to the \textit{generalized-gradient approximation (GGA)} \cite{Perdew1992,Perdew1996}. The underlying idea is to expand the exact DFT potential as a functional of electronic density in a series where the zero-order term is the LDA expression (\ref{e:LDA_functional}) and the first-order term is GGA. Next term contains the Laplacian of density ($\nabla^2 n$) and forms the basis of the meta-GGA method \cite{Tao2003}. The process of increasing the accuracy of DFT by constructing more complex density functionals is usually illustrated by the \textit{Jacob's ladder} (Fig.~\ref{f:DFT_methods}a) where each higher step represents a better approximation to the exact DFT functional. In principle, DFT approximations with higher-order terms should provide more accurate results than LDA. However, higher-order terms have to be constructed such that several properties of the exact DFT functional (so-called sum rules \cite{Perdew2003}) are satisfied.

Even though GGA is built in this way, it does not always improve the theory predictions for the structural and magnetic properties compared to LDA, which probably has to do with error cancellation. For example, while LDA in general underestimates the lattice parameters of most systems, GGA usually overestimates it by a similar amount. This kind of error can be critical when studying systems the functional properties of which are sensitive to structural details, such as ferroelectric where the predicted stability of ferroelectric polarization can change dramatically depending on the DFT approximation \cite{Nishimatsu2010}. Possible ways to improve the DFT description in such cases are i) HSE06 hybrid functional \cite{Heyd2003} and ii) PBEsol functional \cite{Perdew2008}. In case of HSE06, a portion of exact exchange energy (around 25\%) is added to the PBE functional, which also improves the theory prediction for the electronic band gap \cite{Perdew1996b}. In case of PBEsol, the predicted lattice parameter lies in-between the LDA and GGA predictions and is usually very close to the experimental value. This is useful for modelling the mechanical properties or phenomena induced by pressure, because the PBEsol predictions match better the measurements at zero pressure. It should be also noted that the computational complexity of PBEsol is comparable to usual PBE and is much lower than that of HSE06.

In practice, the Kohm-Sham equations (\ref{e:KS_equations}) of DFT are solved in a certain basis, for example, plane waves (Quantum Espresso code \cite{Giannozzi2009,Giannozzi2017}), projector augmented waves (VASP code \cite{Kresse1993,Kresse1996}), linear muffin-tin orbitals (RSPT code \cite{Wills1987,Wills2010}) etc. Moreover, one can differentiate between all-electron and pseudopotential implementations of DFT codes where the core states are either treated on the same footing as the valence electrons or are excluded from equations by working with pseudized wavefunctions which show a smoother behavior near the atomic cores. In general, all-electron codes are considered the most accurate, while the accuracy of pseudopotential codes depends on the quality of pseudopotentials, which are generated based on all-electron results. There can be some differences in DFT theory predictions depending on the particular implementation aspects discussed above, and a detailed analysis can be found in \cite{Lejaeghere2016}.

\subsection{Beyond-DFT approaches}

For systems where electronic correlations play a significant role, it may be required to include further corrections in the DFT functional. Lack of correlations in the original LDA and GGA functionals can, for example, lead to false predictions of metallic behavior or a considerably underestimated value of the band gap (see e.g.~chapter 4 in \cite{Correl12}). This is especially the case for transition-metal and rare-earth compounds with localized \textit{d} or \textit{f} electrons. In these systems, electron interactions correspond to energy scale between $\unit[1-10]{eV}$. In contrast, the exchange splitting in LSDA which produces a finite spin polarization is of the order of $\unit[1]{eV}$ (see for example Table II in \cite{Locht2016}) which corresponds to the Hund's exchange and explains the failure of LSDA (and GGA) for moderately and strongly correlated systems. The DFT+$U$ approach \cite{Anisimov1991,Dudarev1998,Petukhov2003} can be very useful in such cases, since it explicitly introduces additional Coulomb repulsion for chosen states in the spirit of the static mean-field Hubbard model.

The total energy in the DFT+$U$ method contains the usual DFT part (LDA or GGA) and two additional terms, the Coulomb repulsion term and double-counting correction.

\textbf{a) The Couloumb repulsion term} reads (according to \cite{Liechtenstein1995}):
\begin{equation}
    E_\mathrm{corr} = \frac{1}{2} \sum\limits_\gamma (U_{\gamma_1\gamma_3\gamma_2\gamma_4} - U_{\gamma_1\gamma_3\gamma_4\gamma_2}) \hat{n}_{\gamma_1\gamma_2} \hat{n}_{\gamma_3\gamma_4}
    \label{e:Coulomb_term}
\end{equation}

Here, $\hat{n}_{\gamma_1\gamma_2}$ are on-site occupation numbers expressed in the combined orbital and spin space and $U_{\gamma_1\gamma_3\gamma_2\gamma_4}$ are the matrix elements of Coulomb interaction which, in the basis of atomic-like orbitals, can be written in terms of a few parameters (see page 13 of chapter 6 in \cite{Corr11}).

\begin{figure}
{\centering
\includegraphics[width=0.99\textwidth]{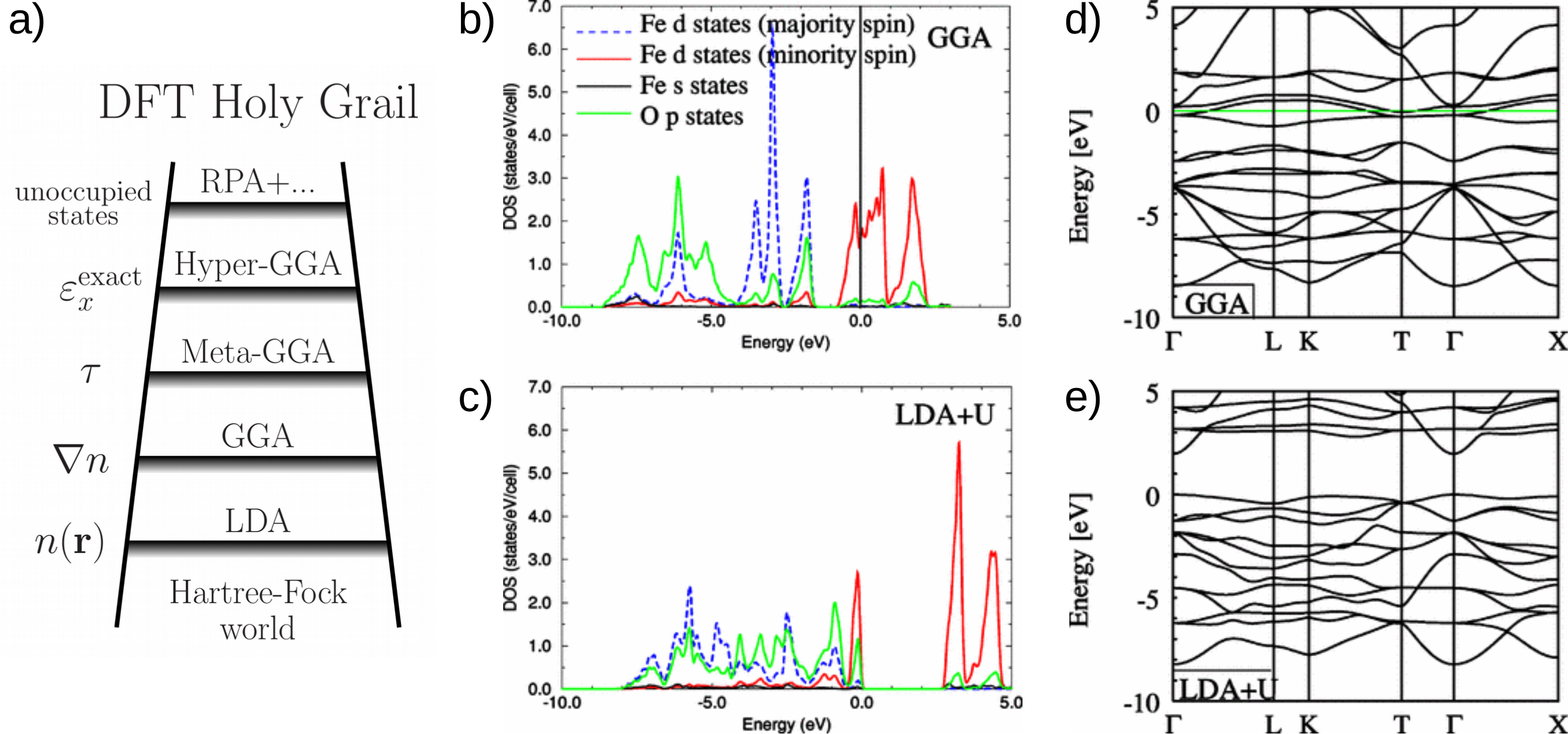}
}
\caption{a) Jacob's ladder of DFT exchange-correlation funtionals with progressing accuracy. Reproduced from thesis \cite{Borisov-thesis} and adapted from the talk ``Basics of DFT'' of K.~Burke and L.~Wagner during the ELK-2011 conference. Comparison of the DFT and DFT+$U$ predictions for b,c) the density of states and the d,e) band structure of the antiferromagnetic Mott insulator FeO (reproduced from \cite{Cococcioni2005}).}
\label{f:DFT_methods}
\end{figure}
In Eqn.~(\ref{e:Coulomb_term}), there are terms that add an energy penalty when two electrons of opposite spin occupy the same orbital and atomic site. This favors half-filled orbitals and leads to finite spin polarization induced by electronic correlations. In case DFT already predicts a magnetic state, the $U$-related term leads to enhancement of magnetic moments and possibly can open or increase the electronic band gap. For example, DFT calculation for antiferromagnetic insulating FeO oxide predicts a metallic system (Fig.~\ref{f:DFT_methods}b), while the DFT+$U$ method correctly predicts an insulating behavior (Fig.~\ref{f:DFT_methods}c) with a band gap of $\unit[2]{eV}$ according to the band structure (Fig.~\ref{f:DFT_methods}d) and increases the Fe moment, bringing it closer to the measured value of $\unit[1.9]{\mu_\mathrm{B}}$. Similar improvement can be achieved for other antiferromagnetic oxides (NiO, CoO etc.) and many further systems (chapter 4 in \cite{Correl12}). The choice of the $U$ parameter can be made based on the agreement between theory and experiment in terms of predicted structural, electronic and magnetic properties or based on independent calculations using \textit{constrained random-phase approximation} \cite{Springer1998} (calculation examples in \cite{Aryasetiawan2004,Aryasetiawan2006,Karlsson2010,Sasioglu2011,Vaugier2012}). Usually, the $U$ parameter is around a few electron volt for moderately correlated systems and can approach $\unit[10]{eV}$ for strongly correlated systems, such as oxides (Table~I in \cite{Anisimov1991}) and rare-earth elements and compounds \cite{Marel1988,Pourovskii2007,Nilsson2013}.

There are also other terms in Eqn.~\ref{e:Coulomb_term} which correspond to the Hund's coupling between different orbitals of the same atomic site and favor parallel alignment of spins on those orbitals. More detailed discussion of these two kinds of interactions will be done in Sec.~\ref{t:spin_models} where we will see how one can derive an effective spin model from the electronic Hamiltonian.

\textbf{b) The double counting term} is necessary, because the approximate exchange-correlation functional in DFT already contains some portion of true electronic correlation and one has to subtract this contribution before adding the Coulomb repulsion term (\ref{e:Coulomb_term}) to avoid double-counting (DC) correlations. There is no unique way of introducing DC corrections and its choice depends on the nature of correlated orbitals. Two widely discussed limiting cases are the fully localized (FLL) and around mean-field (AMF) limits. The FLL double-counting correction is discussed in \cite{Liechtenstein1995}:
\begin{equation}
    E_\mathrm{dc}^\mathrm{FLL} = \frac12 U n(n-1) - \frac12 J \left[ n^\uparrow (n^\uparrow - 1) + n^\downarrow (n^\downarrow - 1) \right]
    \label{e:FFL_DC}
\end{equation}

The around mean-field limit is considered, for example, in \cite{Czyzyk1994,Amadon2008,Bultmark2009}, and its simplified form reads:
\begin{equation}
    E_\mathrm{dc}^\mathrm{AMF} = U n^\uparrow n^\downarrow + \frac12 \left[(n^\uparrow)^2 + (n^\downarrow)^2\right] \frac{2l}{2l+1} (U - J),
    \label{e:AMF_DC}
\end{equation}

where $l$ is the orbital quantum number. In both expressions (\ref{e:FFL_DC}) and (\ref{e:AMF_DC}), the quantities $n^\uparrow$ and $n^\downarrow$ are related to the occupation numbers in the Coulomb term (\ref{e:Coulomb_term}) through the definitions $n^\uparrow = \mathrm{Tr}(\hat{n}^\uparrow_{\gamma_1\gamma_2})$ and $n^\downarrow = \mathrm{Tr}(\hat{n}^\downarrow_{\gamma_1\gamma_2})$, and $n = n^\uparrow + n^\downarrow$.

In general, DFT+$U$ corrections make the exchange-correlation potential orbital-dependent and, in case of FLL (Eqn.~\ref{e:FFL_DC}), favor integer orbital occupations. This leads to shifts of band energies by an amount proportional to the $U$ parameter, such that orbitals with $n>1/2$ are shifted downwards in energy, while orbitals with $n<1/2$ are shifted in the opposite direction (see discussion in \cite{Essenberger2011}).

The DFT+$U$ method outlined above addresses the electronic correlations on the static mean-field level in the spirit of Hartree-Fock approach, where quantum operators are replaced by their average values. A significant improvement over this approximation is offered by the \textit{dynamical mean-field theory} \cite{Georges1996,Kotliar2004,Kotliar2006} which also takes into account the temporal fluctuations of orbital occupations on different atomic sites. In this theory, it is possible to describe a metal-insulator transition induced by electronic correlations even in the paramagnetic phase. This is in contrast to the static DFT+$U$ approach where long-range magnetic order might be necessary in the calculation in order to obtain an insulating behavior. Combination of DFT and DMFT is a powerful method of studying realistic material models and has been extended over the last two decades to include also the non-local correlations effects. Since DFT+DMFT is not in the focus of this review, the interested reader is referred to the reviews \cite{Georges1996,Kotliar2004,Kotliar2006} and book 
\cite{Corr11}, for example.

\section{Atomistic spin models}
\label{t:spin_models}

\subsection{Origin of magnetism}

While we talked a lot about magnetism in electronic correlated systems in the previous section, we did not discuss much its physical origin. This section will focus on this topic in the context of magnetic interactions in general, which can appear in such systems and can lead to long-range magnetic order.

From the point of view of classical physics, it is difficult to understand why some systems can be become magnetic in the presence of external field, not to mention permanent magnets like Fe, Co and Ni which have finite total magnetization even in the absence of external field. The problem with classical description was pointed out by Niels Bohr in 1911 (see publication in collected works \cite{Bohr1911}) and later on by Hendrika Johanna van Leeuwen in 1921 \cite{Leeuwen1921}, and their results are nowadays referred to as \textit{Bohr–Van Leeuwen theorem}, which states that the magnetization vanishes in any classical system placed in external magnetic field. The starting point is the classical Hamiltonian for $N$ interacting electrons (see, for example, discussion in \cite{Kubler2021}):
\begin{equation}
    H = \sum\limits_{i=1}^N \frac{1}{2m}\left( \vec{p}_i - \frac{e}{c}\vec{A}_i \right)^2 + V(\vec{r}_1, \vec{r}_2, \ldots, \vec{r}_N),
    \label{e:classical_Hamiltonian}
\end{equation}

which contains the vector potential $\vec{A}_i \equiv \vec{A}(\vec{r}_i)$ and interactions between all the electrons. When calculating the classical partition function $Z_N$, which involves integration over the phase space of the $\vec{p}_i$ and $\vec{r}_i$ variables, one can substitute variables $\vec{p}\,'_i = \vec{p}_i - \frac{e}{c}\vec{A}_i$ and show that the partition function, in fact, does not depend on the vector potential. Since the average magnetization $M = k_\mathrm{B}T\frac{\partial}{\partial H}\ln Z_N$ is proportional to the derivative of $Z_N$ over magnetic field, the magnetic moment vanishes exactly ($M = 0$).

In order to explain magnetism of electrons in solids, one has to take into account their quantum-mechanical behavior and replace (\ref{e:classical_Hamiltonian}) with:
\begin{equation}
    H = \sum\limits_{i=1}^N \frac{1}{2m}\left( \vec{p}_i - \frac{e}{c}\vec{A}_i \right)^2 + V(\vec{r}_1, \vec{r}_2, \ldots, \vec{r}_N) + \sum\limits_{i=1}^N \left( \mu_\mathrm{B} \vec\sigma \cdot \vec{H}_i + \zeta \vec{l}_i\cdot\vec\sigma \right),
    \label{e:quantum_Hamiltonian}
\end{equation}

where $\vec{p}_i$ and $\vec{r}_i$ are now the momenta and position operators which do not commute. As discussed in \cite{Kubler2021}, this quantum Hamiltonian can be derived as an approximation to Dirac's equations. From the later, one also concludes the existence of electron spin which leads to additional terms proportional to Pauli matrices $\vec\sigma$ in Eqn.~(\ref{e:quantum_Hamiltonian}). The Hamiltonian also contains a coupling between the electron spin and orbital moment $\vec{l}_i$.

\begin{figure}
{\centering
\includegraphics[width=0.99\textwidth]{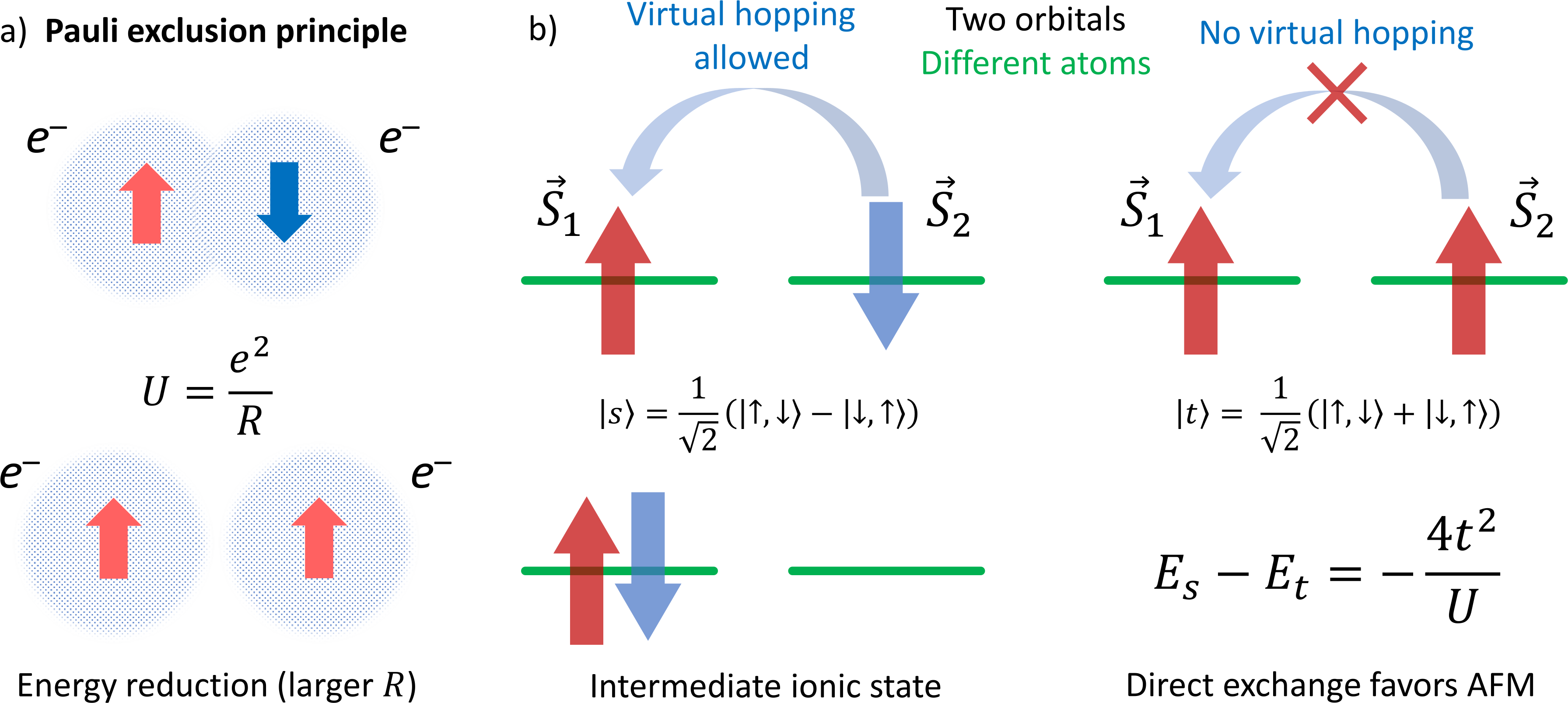}
}
\caption{a) Illustration of the Pauli exclusion principle and exchange interaction for two electrons on the same atomic site. b) Comparison of magnetic energies of antiparallel (singlet state $\left| s \right\rangle$) and parallel (triplet state $\left| t \right\rangle$) spin configurations of two electrons on different sites (each with one orbital). For the antiparallel state the energy is lowered due to existence of electron hopping (with energy $t$) between the sites and intermediate state $\left| \uparrow\downarrow, \_ \right\rangle$ with both electrons on one site. The energy penalty for having both electrons on the same site and orbital equals $U$.}
\label{f:exchange_interaction}
\end{figure}

So far, our discussion addresses an electron system in applied magnetic field. However, even without it some systems are permanent magnets due to the presence of magnetic interactions. It is interesting to see where such interactions can come from, considering that the underlying Hamiltonian (\ref{e:quantum_Hamiltonian}) is \underline{not spin-dependent} for the simple case of $H=0$ and $\zeta = 0$. The well-known answer is that quantum statistics of electronic wavefunction due to the Pauli exclusion principle together with electrostatic Coulomb repulsion between electrons produces the so-called \textit{exchange interaction}. The idea is that two electrons sitting on the same orbital of the same atomic site cannot have the same spin, otherwise they have to be spatially separated by some average distance $R_1$ and occupy different orbitals. This configuration has, in fact, lower Coulomb energy $e^2/R_1$ compared to the case of two electrons with opposite spin, which can be in the same orbital and, accordingly, at a smaller average distance from each other $R_2 < R_1$ (Fig.~\ref{f:exchange_interaction}a). Because of this, the energy of the system becomes \underline{spin-dependent} and can be effectively described by the exchange interaction which favors a ferromagnetic alignment of spins on the same site, known as the \textit{Hund's coupling}.

\subsection{Direct exchange}

Let us extend this consideration onto the case of two electrons occupying two orbitals, each on a different site (Fig.~\ref{f:exchange_interaction}b). This situation is quite different, because now electrons can hope between the two sites leading to six possible electronic configurations: $\left| \uparrow\downarrow,\_ \right\rangle$, $\left| \_,\uparrow\downarrow \right\rangle$, $\left| \uparrow, \downarrow \right\rangle$, $\left| \downarrow, \uparrow \right\rangle$, $\left| \uparrow, \uparrow \right\rangle$, and $\left| \downarrow, \downarrow \right\rangle$, meaning that either both sites are half-occupied or one of them is empty, while the two electron spins can be parallel or antiparallel (detailed discussion of this model is found, for example, in \cite{Correl12}, chapter 7 by Erik Koch). It is more illustrative to write the Hamiltonian for this problem in the second quantization, instead of the first-quantization adopted in Eqn.~(\ref{e:quantum_Hamiltonian}):
\begin{equation}
    H = t\sum\limits_{\sigma=\uparrow,\downarrow} \hat{c}^+_{2\sigma} \hat{c}^{}_{1\sigma} + t\sum\limits_{\sigma=\uparrow,\downarrow} \hat{c}^+_{1\sigma} \hat{c}^{}_{2\sigma} + U\, \hat{c}^+_{1\uparrow} \hat{c}^{}_{1\uparrow} \hat{c}^+_{1\downarrow} \hat{c}^{}_{1\downarrow} + U\, \hat{c}^+_{2\uparrow} \hat{c}^{}_{2\uparrow} \hat{c}^+_{2\downarrow} \hat{c}^{}_{2\downarrow}
    \label{e:twosites_2nd_quant}
\end{equation}

The essential components of this model are i) electron hopping with energy $t$ and ii) Coulomb repulsion energy $U$ between two electrons on the same orbital/site. Diagonalization of this Hamiltonian in the basis of six two-electron states mentioned above leads e.g.~to a triplet eigenstate represented by $\left| \uparrow, \uparrow \right\rangle$, $\left| \downarrow, \downarrow \right\rangle$ and superposition $\frac{1}{\sqrt{2}}\left( \left| \uparrow, \downarrow \right\rangle + \left| \downarrow, \uparrow \right\rangle \right)$ and a singlet state $\frac{1}{\sqrt{2}}\left( \left| \uparrow, \downarrow \right\rangle - \left| \downarrow, \uparrow \right\rangle \right)$ which is lower in energy by $\frac{4t^2}{U}$ (in the leading order of perturbation theory). Thus, in contrast to \textit{Hund's coupling}, \textit{direct exchange} between two sites favors an antiferromagnetic spin alignment.

If we want to build a model of this system with only states that have a fixed spin $S=\frac{1}{2}$ on each site, then we disregard the intermediate states with double occupancy of one of the sites and downfold the original Hamiltonian onto the subspace of states $\left| \uparrow, \downarrow \right\rangle$, $\left| \downarrow, \uparrow \right\rangle$, $\left| \uparrow, \uparrow \right\rangle$, and $\left| \downarrow, \downarrow \right\rangle$, as shown in detail in \cite{Correl12}. After rewriting the resulting second-quantization expression in terms of spin operators, this procedure leads to an \textit{effective spin Hamiltonian}:
\begin{equation}
    H_\mathrm{spin} = \frac{2t^2}{U} \left( \vec{S}_1 \cdot \vec{S}_2 - \frac{n_1 n_2}{4} \right) = \frac{2t^2}{U} \left( \vec{S}_1 \cdot \vec{S}_2 \right) + \mathrm{const}
    \label{e:two_site_spin_model}
\end{equation}

This downfolding from the full electronic space to a spin subspace filters out the high-energy states with double occupancy but still takes into account their effect on other states leading to effective spin-spin interactions, so the effective Hamiltonian (\ref{e:two_site_spin_model}) is spin-dependent while the original one (\ref{e:twosites_2nd_quant}) is spin-independent. The resulting spin model (Eqn.~\ref{e:two_site_spin_model}) has the form of the \textit{Heisenberg model}, which was derived for the first time by Werner Heisenberg in 1928 \cite{Heisenberg1928} for quantum $S=1/2$ systems, followed by numerous discussions in the literature, for example, in the context of single-band Hubbard model for extended systems \cite{Takahashi1977,MacDonald1988}.

\subsection{Beyond-Heisenberg models}

\begin{figure}
{\centering
\includegraphics[width=0.99\textwidth]{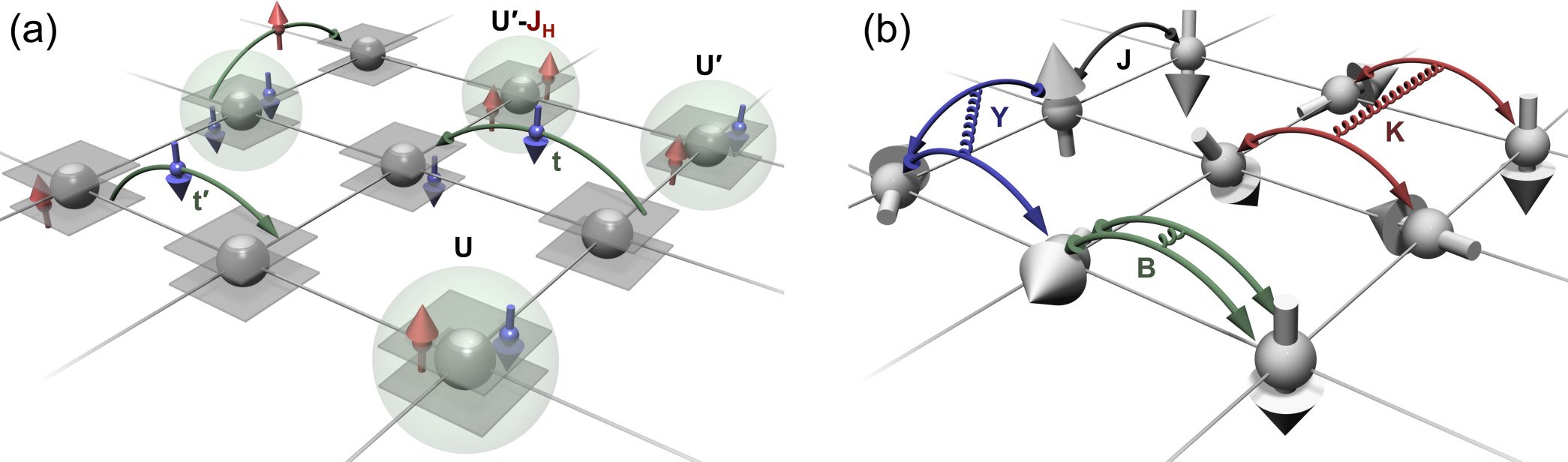}
}
\caption{a) Electronic and b) effective spin models for a periodic crystal lattice (reproduced from \cite{Hoffmann2020}). a) The electronic Hamiltonian  corresponds to the Hubbard model and contains electron hopping terms, Coulomb interactions between electrons on the same site in the same orbital ($U$) or different orbitals ($U'$) as well as the Hund's coupling $J_\mathrm{H}$. b) The spin model derived from the Hubbard model includes, for example, the bilinear two-site ($J$) and four-site ($K$) interactions, three-spin exchange ($Y$) and the biquadratic interaction $B$.}
\label{f:effective_spin_model}
\end{figure}
The procedure of deriving effective spin model from the electronic Hamiltonian described above is general and can be applied to a variety of electronic systems. Didactic discussion of different aspects of this procedure from the point of view of perturbation theory is given in \cite{Hoffmann2020} (accompanying software is provided in \cite{Hoffmann2023}), where strongly correlated electrons on a square lattice, for example, are considered and the electron hopping between the neighboring sites ($t\ll U$) is treated as a perturbation. For $S=1/2$ systems with Hamiltonian like Eqn.~(\ref{e:two_site_spin_model}) generalized to periodic crystal lattices, it can be shown that the effective spin model contains two kind of terms:
\begin{equation}
    H_\mathrm{spin} = -\sum\limits_{ij} J_{ij}(\vec{S}_i\cdot\vec{S}_j) - \sum\limits_{ijkl} K_{ijkl} (\vec{S}_i\cdot\vec{S}_j)(\vec{S}_k\cdot\vec{S}_l),
    \label{e:spin12_model}
\end{equation}

where the summations run for $i\neq j\neq k\neq l$. The first term here corresponds again to the Heisenberg model, while the additional term represents a four-spin interaction. So already in this simple case one obtains beyond-Heisenberg terms in the spin model which describe more complex excitations of the electronic system and include multi-site electron hopping processes. Different orders of the perturbation theory for $S=1/2$ systems bring corrections to the exchange parameters $J_{ij}$ and $K_{ijkl}$ in Eqn.~(\ref{e:spin12_model}) but do not produce any further types of spin-spin interactions.

Important to note is that the spin model (\ref{e:spin12_model}) derived from the electronic Hamiltonian is a \textit{quantum spin model} with $S=1/2$, so $\vec{S}_1$ and $\vec{S}_2$ are operators with commutation relationships common for angular momentum operators.

\label{t:higher_order_spin}
For higher-spin systems $S\geq 1$, additional contributions to the spin model are possible, such as the biquadratic and three-spin interactions for $S=1$ case:
\begin{equation}
    H_\mathrm{bq} = -\sum\limits_{ij} B_{ij} (\vec{S}_i\cdot\vec{S}_j)^2
    \label{e:biquadratic_exchange}
\end{equation}
\begin{equation}
    H_\mathrm{3-spin} = -\sum\limits_{ijk} Y_{ijk} (\vec{S}_i\cdot\vec{S}_j)(\vec{S}_i\cdot\vec{S}_k)
    \label{e:3spin_exchange}
\end{equation}

Due to the quantum nature of spin Hamiltonian, biquadratic exchange does not appear for $S=1/2$ systems, because it can be reduced to a usual bilinear Heisenberg interaction due to the properties of spin-ladder operators (see for example \textbf{[refs]}). Similarly, on-site anisotropy of the form $K_U (\vec{S}_i\cdot\vec{e})^2$ vanishes exactly for $S=1/2$ systems, since this terms is equivalent to applying twice the ladder operators $\hat{S}^+$ and $\hat{S}^-$ to $\left| m_z=-1/2 \right\rangle$ and $\left| m_z=+1/2 \right\rangle$ states, respectively. In general, terms up to order $(\vec{S}_i\cdot\vec{S}_j)^{2S}$ are allowed in the quantum spin Hamiltonian for spin number $S$.

Adding spin-orbit coupling to the original electronic Hamiltonian in the form $\lambda (\vec{L}_i\cdot\vec{S}_i)$ and including it in the perturbation theory expansion would lead to new types of magnetic interactions in the effective spin model, such as the \textit{Dzyaloshinskii-Moriya} (DM) interaction $H_\mathrm{DM}=\vec{D}_{ij}\cdot(\vec{S}_i\times\vec{S}_j)$  and its higher-order variants, e.g.~$H^{(2)}_
\mathrm{DM}=\vec{D}'_{ij}\cdot(\vec{S}_i\times\vec{S}_j) (\vec{S}_i\cdot\vec{S}_j)$. The bilinear DM interaction $H_\mathrm{DM}$ can exist in magnetic systems where the inversion symmetry is broken for bonds connecting magnetic moments. Such symmetry breaking can be especially strong near surfaces or interfaces, and well-known examples are transition metal multilayers, e.g.~Pd/Fe/Ir(111)~\cite{Romming2013}, Pd/Co/Pd~\cite{Pollard2017,Brandao2019,Wei2021,Carvalho2023}, Pt/Co/Ta \cite{Wang2019}, Ir/Fe/Co/Pt \cite{Soumyanarayanan2017} and exchange-biased multilayers \cite{Rana2020}, to name just a few. The DM interaction favors non-collinear magnetic orders with a certain chirality defined by the DM vectors $\vec{D}_{ij}$ and, historically, was proposed in \cite{Dzyaloshinsky1958,Moriya1960} to explain the phenomenon of canted magnetism in $\alpha$-Fe$_2$O$_3$, MnCO$_3$ and CoCO$_3$. For certain crystal symmetries, it can even stabilize topogical magnetic textures with a size below a hundred nanometers (for reviews see \cite{Kanazawa2017,Bihlmayer2018,EverschorSitte2018,Goebel2021,Zhang2023}). More details about the requirements on symmetry are given in Sec.~\ref{t:micromagnetics} of this review.

\subsection{LKAG first-principles approach}
\label{t:LKAG}

While the approach described above provides, in principle, the values of magnetic interactions, these are rather approximate, because they are obtained within the perturbation theory where $t/U$ is assumed to be a small parameter ($t$~-- electron hopping, $U$~-- strength of electronic correlations). More accurate approach was suggested in the LKAG paper~\cite{LKAG1987} where the idea is to calculate the energy change $\Delta E_{ij}$ due to a small perturbation of the reference magnetic state. More specifically, this perturbation includes an infinitesimal canting of two spins on sites $i$ and $j$ (Fig.~\ref{f:multiscale_approach}b). In the non-relativistic case, the energy change $\Delta E_{ij}$ is then proportional to the corresponding Heisenberg exchange parameter $J_{ij}$. The later can be calculated from perturbation theory, but this time the advantage is that the canting angle is, indeed, a small parameter, in contrast to $t/U$, so the LKAG approach is more accurate and is applicable, in principle, to any magnetic system with intrinsic moments. Using the language of Green functions, the Heisenberg exchange in the LKAG approach can be expressed as follows:
\begin{equation}
    J_{ij} = \frac{1}{4\pi} \mathrm{Im} \int\limits_{-\infty}^{E_\mathrm{F}} \mathrm{Tr} [\hat\Delta_i \hat G_{ij}^{\uparrow}(\varepsilon) \hat\Delta_j \hat G_{ji}^{\downarrow}(\varepsilon) ]
	\label{e:LKAG_nonrel}
\end{equation}

As we see, the Heisenberg exchange depends on the electron Green function for two sites ($i$ and $j$) which is reminiscent of electron hopping parameter $t$ discussed on previous pages, and is proportional to the spin splitting $\Delta$ of electronic states on these sites. Both quantities ($\hat{G}_{ij}$ and $\hat{\Delta}$) contain the information on the electronic structure, making the LKAG approach material-specific, and are reprensented by matrices in orbital space. Correspondingly, the trace in Eqn.~(\ref{e:LKAG_nonrel}) is equivalent to summing over different orbitals. However, it is possible to analyze orbital-resolved contributions by skipping the trace operator and internal matrix multiplications, leading to expression \cite{Kvashnin2016,Cardias2017}:
\begin{equation}
    J_{ij}^{m_1 m_2} = \frac{T}{4}\sum_n \Delta_i^{m_1}  G_{ij}^{\uparrow m_1 m_2} (i\omega_n) \Delta_j^{m_2} G_{ji}^{\downarrow m_1 m_2}(i\omega_n),
	\label{e:LKAG_nonrel_orbital}
\end{equation}

which is written now in terms of summation over Matsubara frequencies $\omega_n$.

The LKAG approach is a widely used and efficient method for calculating magnetic interactions in electronic systems, both periodic and non-periodic. It allows to study the long-range character of interactions while using the minimal chemical unit cell, in contrast to the total energy mapping method which always requires constructing large supercells leading to high computational costs. Detailed discussion of magnetic interactions, ways to calculate them and examples for solid state systems can be found in the comprehensive review \cite{Szilva2022}. In the original LKAG paper~\cite{LKAG1987}, the authors also discussed how to estimate the Curie temperature and spin stiffness, that can be directly compared to experiment, and calculations for Fe, for example, showed a good agreement between theory and experiment. Furthermore, the exchange parameters obtained in the LKAG approach correspond to energy variations at zero temperature and, for that reason, naturally describe the magnon excitations with a high accuracy (for a review, see \cite{Etz2015}).

\begin{figure}
{\centering
\includegraphics[width=0.99\textwidth]{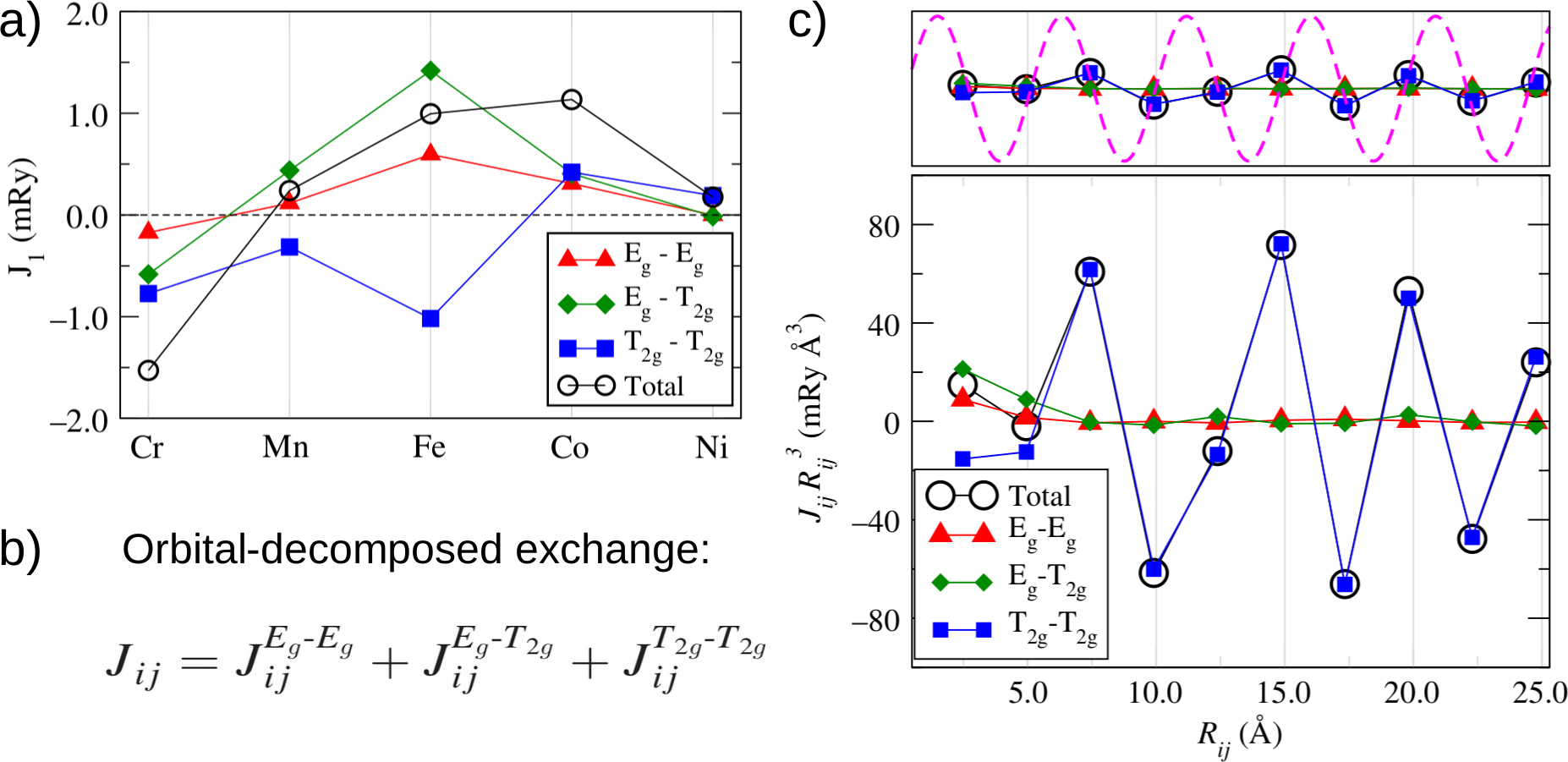}
}
\caption{a) Orbital decomposition of Heisenberg exchange in magnetic transition metals; b) different contributions from $e_g$ and $t_{2g}$ orbitals. c) RKKY character of Heisenberg interaction between the $t_{2g}$ orbitals in Fe \textit{bcc} and short-range character of interactions involving the $e_g$ orbitals. Reproduced from \cite{Kvashnin2016}.}
\label{f:Jij_RKKY_in_metals}
\end{figure}

Regarding the long-range character of Heisenberg interaction, the LKAG approach has been useful for analyzing the mechanism of magnetic exchange in transition metals Cr, Mn, Fe, Co, and Ni. In \cite{Kvashnin2016,Cardias2017}, orbital decomposition of Heisenberg exchange revealed oscillations of $t_{2g}$-related contributions as $\sim\!\!\sin(k_\mathrm{F}r)/r^3$ according to the Ruderman–Kittel–Kasuya–Yosida mechanism \cite{Ruderman1954,Kasuya1956,Yosida1957} and short-range character of $e_g$-related contributions (Fig.~\ref{f:Jij_RKKY_in_metals}). The $t_{2g}$ contributions are affected by the properties of the Fermi surface; for example, the Fermi wavevector $k_\mathrm{F}$ determines the oscillation period of the Heisenberg exchange over distance, while the $e_g$ contributions depend on the electron hopping parameter resembling the double exchange. Such analysis makes the connection between the electronic properties and magnetic interactions clearer and helps to understand and predict how the magnetic interactions change in response to structural and chemical variations.

The LKAG formula (\ref{e:LKAG_nonrel}) has been extended for the relativistic case to provide information on the Dzyaloshinskii-Moriya interaction and for correlated systems \cite{Udvardi2003,Ebert2009,Secchi2015,Kvashnin2020}, such that, for example, the $J_{xy}$ component of the generalized exchange tensor looks like this:
\begin{equation}
    J_{ij}^{xy} = -\frac{T}{4} \sum\limits_n \mathrm{Tr} \Big( [\hat{H}_i + \hat{\Sigma}_i(i\omega_n), \hat\sigma^y] G_{ij}(i\omega_n) [\hat{H}_j + \hat{\Sigma}_j(i\omega_n), \hat\sigma^x] G_{ji}(i\omega_n) \Big)
\end{equation}

The important difference here is the presence of frequency-dependent self-energy $\Sigma_i(i\omega_n)$, which can be calculated within the dynamical mean-field theory approach \cite{Georges1996,Kotliar2004,Kotliar2006}. Furthermore, the Pauli matrices $\hat{\sigma}^x$ and $\hat{\sigma}^y$ describe the non-collinear character of the spin configuration due to the canting of the considered spins along the $x$- and $y$-directions. From the off-diagonal components like $J_{ij}^{xy}$ one can calculate the Dzyaloshinskii-Moriya interaction $\vec{D}_{ij}$ and symmetric anisotropic exchange $\hat{\Gamma}_{ij}$, while the Heisenberg exchange is obtained from the diagonal components $J_{ij}^{\alpha\alpha}$, so that the full exchange tensor is written as follows (see discussion in \cite{Borisov2022}):
\begin{equation}
    \hat{J}_{ij} = 
    \left(
    \begin{array}{ccc}
        J^{xx}_{ij} & \Gamma^{xy}_{ij}+D^z_{ij} & \Gamma^{xz}_{ij}-D^y_{ij} \\[5pt]
        \Gamma^{xy}_{ij}-D^z_{ij} & J^{yy}_{ij} & \Gamma^{yz}_{ij}+D^x_{ij} \\[5pt]
        \Gamma^{xz}_{ij}+D^y_{ij} & \Gamma^{yz}_{ij}-D^x_{ij} & J^{zz}_{ij}
    \end{array}
    \right) \label{e:J_matrix}
\end{equation}

and the effective spin model is represented by the generalized Heisenberg model:
\begin{equation}
    H = -\sum\limits_{i\neq j} S_i^\alpha \hat J_{ij}^{\alpha\beta} S_j^\beta, \hspace{10pt} \alpha,\beta=x,y,z,
    \label{e:general_Heisenberg_model}
\end{equation}

where $\vec{S}_i$ are unit vectors showing the direction of the local spin axis on site $i$. This means that the $J_{ij}$ parameters calculated in the LKAG formalism already contain the magnitude of the magnetic moments on individual sites. This should be kept in mind when comparing these $J_{ij}$ values to the results obtained using other methods.

It should be noted that the magnetic interactions in the LKAG approach are evaluated for a certain spin configuration called \textit{reference state}. Very often it makes sense to use the lowest energy spin configuration as the reference state or a magnetic state which is close to the ground state, if the later is too complicated to treat within DFT (e.g. large-wavelength spin spiral). If the reference state is non-collinear and different from the ground state, then it is necessary to apply a constraining field to stabilize such a configuration. However, in that case, for DFT-based calculations of exchange interactions one has to bear in mind that the constraining field is different from the gradient of the total magnetic energy obtained from DFT \cite{Streib2020}. This is reflected in the way that the electronic system is mapped onto an effective spin model, as elaborated upon in recent works \cite{Streib2021,Streib2022}.

\section{Micromagnetic description}
\label{t:micromagnetics}

\subsection{Basic equations}

Although effective spin models, discussed in the previous section, are a powerful tool for studying the time-dependent behavior and temperature-dependent properties of magnetic systems based on atomistic spin dynamics equations (ASD, for details see Sec.~\ref{t:LLG_equations}), there are limits to the size of spin systems that can be modelled on modern computing architectures. For large systems, the simulation time scales almost linearly with the total number of spins $N_\mathrm{spins}$ and, in practice, such atomistic simulations are feasible for up to a hundred million spins. For 2D systems described by $(n\times n)$ supercells ($N_\mathrm{spins} \sim n^2$), this gives an excellent opportunity to study complex spin textures, such as domain walls, spin spirals, skyrmions etc.~on the length scale of up to a few hundred nanometers. However, for 3D systems the total number of spins grows faster with the system dimensions ($N_\mathrm{spins} \sim n^3$), which reduces significantly the size of systems that can be modelled atomistically, in terms of simulation time and required memory.

In order to address the magnetic properties of larger systems, one can switch to the micromagnetic description where the underlying atomic structure is completely disregarded in the simulation and the magnetization is treated as a continuous vector field $\vec{m}(\vec{r})$ defined at each point of space, not just at atomic sites. Of course, in the actual simulation $\vec{m}(\vec{r})$ is discretized on e.g.~a 3D grid ($n\times n\times n$), and the simplification here, compared to ASD simulations, is related to the fact that each of the $n^3$ micromagnetic elementary cells usually covers around a few nanometers of the system in each spatial direction. Thus, it requires less micromagnetic cells than actual atomic spins to model a system of a given size.

Further below, the basic equations and definitions of the micromagnetic approach are derived in a didactic way, followed by details about numerics and concrete examples from literature.

\vspace{5pt}
\textbf{Heisenberg interactions.} Let us first derive the micromagnetic model corresponding to the Heisenberg exchange (first term in Eqn.~\ref{e:Heisenberg_model}). Following the methodology discussed in Ref.~\cite{Poluektov2018}, one can start from the atomistic picture of spins on sites $i$ and $j$ and write $\vec{S}_j = \vec{S}_i + (\vec{S}_j - \vec{S}_i)$. The difference in the brackets by definition is equal to $\vec{m}(\vec{r}_j) - \vec{m}(\vec{r}_i)$, since the magnetization in different regions is described by the micromagnetic function $\vec{m}(\vec{r})$ which can be expanded in a Taylor series:
\begin{equation}
    \vec{m}(\vec{r}_j) - \vec{m}(\vec{r}_i) \approx (\vec{R}_{ij}\cdot\vec{\nabla})\vec{m} + \frac12 (\vec{R}_{ij}\cdot\vec{\nabla})^2\vec{m}
\end{equation}

Here, $\vec{R}_{ij}$ is the vector in real space connecting spins on sites $i$ and $j$, and $\vec{m} \equiv \vec{m}(\vec{r_i})$.

The magnetic energy for the Heisenberg exchange can be written now as follows:
\begin{equation}
    \varepsilon_\mathrm{H} = \sum\limits_{\left\langle ij\right\rangle} J_{ij} (\vec{S}_i \cdot \vec{S}_j) \rightarrow \sum\limits_{\left\langle ij\right\rangle} J_{ij} \,\vec{m}\cdot \left( \vec{m} + (\vec{R}_{ij}\cdot\nabla)\vec{m} + \frac12 (\vec{R}_{ij}\cdot\nabla)^2\vec{m} \right).
    \label{e:Heisenberg_micromagnetic}
\end{equation}

After writing out the expression in the brackets, we see that the first term is a constant energy contribution, since it is proportional to $\vec{m}^2 = 1$ (constant length of magnetic moment vectors is assumed). The second term with a 1$^{\mathrm{st}}$-order derivative can be rewritten:
\begin{equation}
    \vec{m}\cdot (\vec{R}_{ij}\cdot\nabla)\vec{m} \equiv m_\alpha R_{ij}^\beta \nabla_\beta m_\alpha = R_{ij}^\beta \nabla_\beta \left( m_\alpha^2/2 \right).
\end{equation}

and is equal to zero, since $m_\alpha m_\alpha = 1$.

For that reason, the first non-vanishing term in Eqn.~(\ref{e:Heisenberg_micromagnetic}) reads:
\begin{equation}
    \varepsilon_\mathrm{H} = \frac12 \sum\limits_{\left\langle ij\right\rangle} J_{ij} \,\vec{m}\cdot (\vec{R}_{ij}\cdot\nabla)^2\vec{m} = \frac12 \sum\limits_{\left\langle ij\right\rangle} J_{ij} R_{ij}^\alpha R_{ij}^\beta \, m_\gamma \nabla_\alpha \nabla_\beta m_\gamma.
\end{equation}

Usually, only the diagonal terms ($\alpha = \beta$) are considered in micromagnetics and, for equal diagonal elements $A_{xx} = A_{yy} = A_{zz} = A$ (e.g.~in cubic systems), the magnetic energy density is written as follows:
\begin{equation}
    \varepsilon_\mathrm{H} = A \left[ m_x \nabla^2 m_x + m_y \nabla^2 m_y + m_z \nabla^2 m_z \right]
    \label{e:micromagnetic_functional_Heisenberg}
\end{equation}

The energy is proportional to the so-called spin stiffness $A$:
\begin{equation}
  A = \frac{1}{2} \sum_{\left\langle ij\right\rangle} J_{ij} R^2_{ij},
  \label{e:spin_stiffness}
\end{equation}

which characterizes the overall strength of magnetism in the system. For example, for ferromagnets ($A>0$) one expects that the Curie temperature increases with the spin stiffness $A$. We also see that the whole complexity of magnetic interactions $J_{ij}$ between different spin neighbors is boiled down to just one number $A$, which reflects the simplifying character of the micromagnetic approach. However, the spin stiffness $A$ is obtained from atomistic interactions $J_{ij}$ which are calculated for a given crystal structure and chemical composition from the first principles of quantum mechanics. This means that the micromagnetic model is also obtained from first principles and reflects the properties of a concrete material, which is a great advantage of the \textit{multiscale approach}.

\vspace{5pt}
\textbf{Dzyaloshinskii-Moriya interactions.} Now let us derive the micromagnetic energy density $\varepsilon_\mathrm{DM}$ originating from the DM interactions, following the same methodology as outlined above for the Heisenberg exchange. The atomistic spin model (second term in Eqn.~(\ref{e:Heisenberg_model})) is the starting point:
\begin{equation}
    \varepsilon_\mathrm{DM} = -\sum\limits_{\left\langle ij\right\rangle} \vec{D}_{ij} \cdot (\vec{S}_i \times \vec{S}_j)
    \label{e:DMI_atomistic}
\end{equation}

Again, we replace $\vec{S}_i$ with the micromagnetic function $\vec{m}\equiv\vec{m}(\vec{r})$ and expand $\vec{S}_j$ up to the 1$^\mathrm{st}$-order term: $\vec{m}+(\vec{R}_{ij}\cdot\vec{\nabla})\vec{m}$, where $\vec{R}_{ij}$ is the distance vector between spins $i$ and $j$. Substituting this into 
Eqn.~(\ref{e:DMI_atomistic}) and noticing that $\vec{m}\times\vec{m} = 0$ leads to a micromagnetic energy density:
\begin{equation}
    \varepsilon_\mathrm{DM} = -\sum\limits_{\left\langle ij\right\rangle} \vec{D}_{ij} \cdot (\vec{m}\times (\vec{R}_{ij}\cdot\vec{\nabla})\vec{m}) = +\vec{m}\cdot \left[ \sum\limits_{\left\langle ij\right\rangle}\vec{D}_{ij} (\vec{R}_{ij}\cdot\vec{\nabla})\right]\times \vec{m},
    \label{e:DM_micromagnetic_derivation}
\end{equation}

where one can define the \textit{spiralization matrix} $\hat{D} \equiv D_{\alpha\beta}\:(\alpha,\beta=x,y,z)$ as
\begin{equation}
  D_{\alpha\beta} = \sum_{\left\langle ij\right\rangle} D_{ij}^\alpha R_{ij}^\beta \hspace{5pt}\rightarrow \hspace{5pt}\varepsilon_\mathrm{DM} = m_\gamma \,\varepsilon_{\gamma\alpha\delta} D_{\alpha\beta}\nabla_\beta \,m_\delta
  \label{e:spiralization_matrix}
\end{equation}

The spiralization matrix can have nine non-zero components and reflects the crystal and magnetic symmetries. For systems with chiral crystal structure, such as cubic B20 compounds MnSi \cite{Muehlbauer2009}, FeGe \cite{Lundgren1968,Wappling1968,Lebech1989,Yu2011} and Fe$_{1-x}$Co$_{x}$Si \cite{Yu2010}, the spiralization matrix is diagonal ($D_{xx} = D_{yy} = D_{zz} = D$) and this scenario is referred to as the so-called \textit{bulk DM interaction}. By substituting the matrix elements $D_{\alpha\beta} = D\,\delta_{\alpha\beta}$ one can obtain the micromagnetic energy:
\begin{equation}
    \varepsilon_\mathrm{bulk} = (m_x, m_y, m_z)\cdot\left|
      \begin{array}{ccc}
          \vec{e}_x & \vec{e}_y & \vec{e}_z \\[3pt]
          D\,\frac{\partial}{\partial x} & D\,\frac{\partial}{\partial y} & D\,\frac{\partial}{\partial z} \\[3pt]
          m_x & m_y & m_z
      \end{array}
    \right| = D\,\vec{m}\cdot(\vec{\nabla}\times\vec{m})
    \label{e:bulk_DMI}
\end{equation}

\begin{figure}
{\centering
\includegraphics[width=0.99\textwidth]{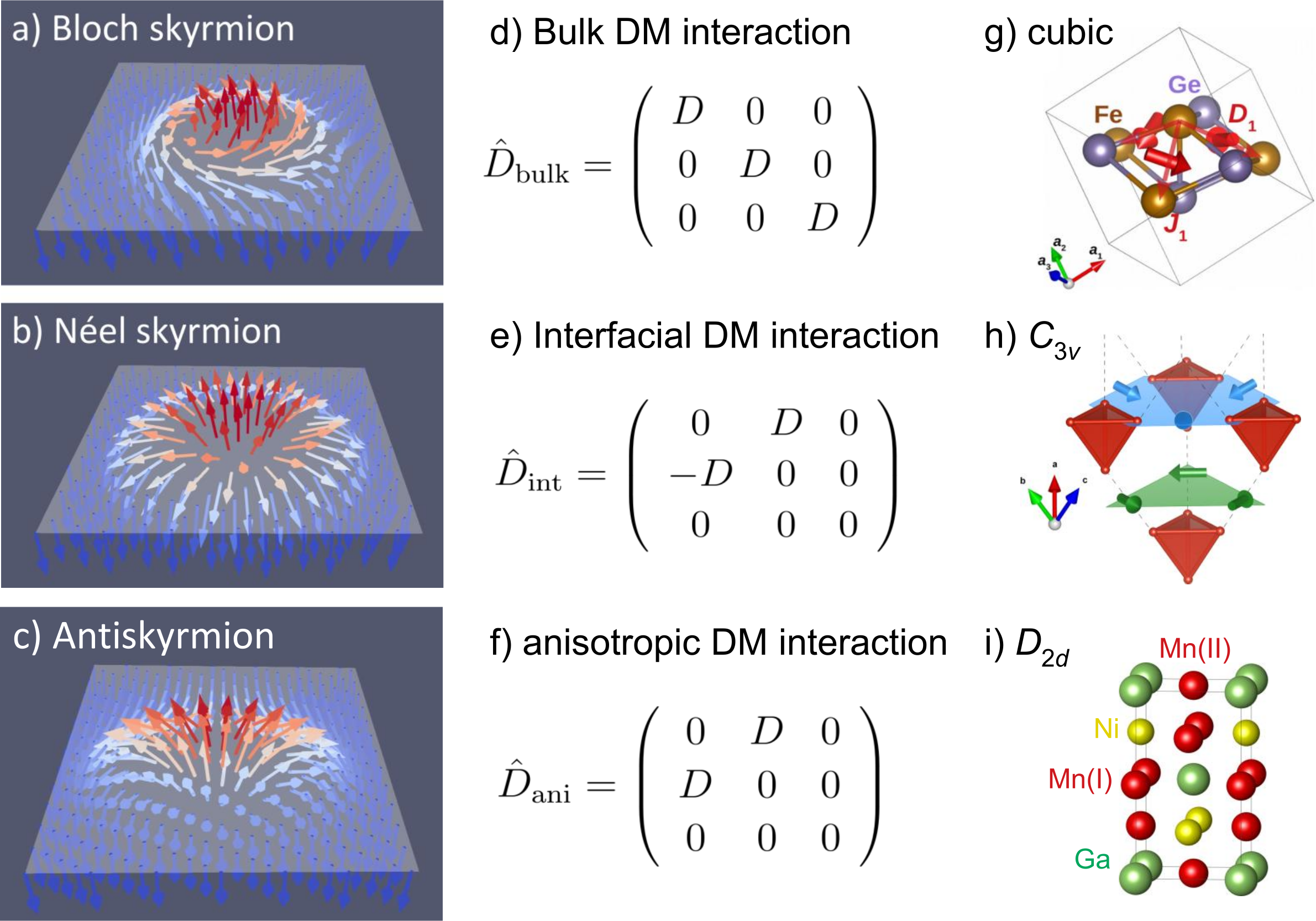}
}
\caption{Schematic illustration of a) Bloch and b) N\'eel skyrmions and c) antiskyrmion that are stabilized by the d) bulk, e) interfacial and f) anisotropic Dzyaloshinskii-Moriya (DM) interactions (the corresponding DM matrices are shown as well). Examples of systems with such interactions are g) B20 compound FeGe, h) multiferroic lacunar spinel GaV$_4$S$_8$ and i) Heusler compound Mn$_2$NiGa. The relevant crystal structures and symmetries of each compound class are shown as well. Figures a) and b) are reproduced from \cite{Borisov2023b}, figure g)~-- from \cite{Borisov2021}, and figure h)~-- from \cite{Borisov2023a}}
\label{f:types_of_DMI}
\end{figure}

This type of DMI favors the formation of \textit{Bloch skyrmions} (Fig.~\ref{f:types_of_DMI}a), which are localized magnetic objects with topologically non-trivial winding of atomic spins (further discussion in Sec.~\ref{t:magn_predictions}).

Different scenario takes place for transition metal multilayers, such as Pd/Fe/Ir(111) \cite{Romming2013} and Pd/Co/Pd \cite{Pollard2017,Brandao2019,Wei2021,Carvalho2023}, which have the $C_{3\nu}$ crystal symmetry. In that case, the symmetry of the DM interaction vectors leads to non-zero off-diagonal elements $D_{xy} = -D_{yx} = D$, while other elements are zero. This scenario corresponds to the so-called \textit{interfacial DM interaction} with a micromagnetic energy which stabilizes \textit{N\'eel skyrmions} (Fig.~\ref{f:types_of_DMI}b):
\begin{equation}
    \varepsilon_\mathrm{int} = (m_x, m_y, m_z)\cdot\left|
      \begin{array}{ccc}
          \vec{e}_x & \vec{e}_y & \vec{e}_z \\[3pt]
          D\,\frac{\partial}{\partial y} & -D\,\frac{\partial}{\partial x} & 0 \\[3pt]
          m_x & m_y & m_z
      \end{array}
    \right|
    \label{e:DM_energy_1}
\end{equation}
\begin{equation}
    \varepsilon_\mathrm{int} = -D \left[ m_x \frac{\partial m_z}{\partial x} - m_z \frac{\partial m_x}{\partial x} + m_y \frac{\partial m_z}{\partial y} - m_z \frac{\partial m_y}{\partial y} \right].
    \label{e:interfacial_DMI}
\end{equation}

This result agrees with the Lifshitz invariants expected for the $C_{3\nu}$ crystal symmetry, as discussed, for example, in~\cite{Bogdanov2001} (Eqn.~8), \cite{Bogdanov2002} (Eqn.~6) and \cite{Ado2020} (Table~I). The same micromagnetic expression is obtained also for polar crystals, such as lacunar spinels GaV$_4$S$_8$ \cite{Borisov2023a} and GaV$_4$Se$_8$ \cite{Zhang2017}, which at low temperature both have the $C_{3\nu}$ symmetry.

There is another class of materials, Heusler compounds, which have $D_{2d}$ crystal symmetry. Magnetic compounds of this type, Mn$_{1.4}$Pt$_{0.9}$Pd$_{0.1}$Sn \cite{Peng2020,Jena2020} and Mn$_2$NiGa \cite{Sen2019}, show the so-called \textit{anisotropic DM interaction} characterized by a DM matrix with the only non-vanishing components $D_{xy} = D_{yx} = D$. We notice here that both off-diagonal elements have the same sign, as opposed to the interfacial DMI discussed above. This kind of structure of DMI leads to the micromagnetic energy (notice sign changes compared to Eqn.~\ref{e:interfacial_DMI}):
\begin{equation}
    \varepsilon_\mathrm{ani} = -D \left[ -m_x \frac{\partial m_z}{\partial x} + m_z \frac{\partial m_x}{\partial x} + m_y \frac{\partial m_z}{\partial y} - m_z \frac{\partial m_y}{\partial y} \right],
    \label{e:anisotropic_DMI}
\end{equation}

which favors \textit{antiskyrmions} (Fig.~\ref{f:types_of_DMI}c). In terms of the spin winding, the later are in-between the N\'eel and Bloch skyrmions, because the spins are winding perpendicular to the radius vector from the center of the antiskyrmion in one direction and parallel to it in other directions. Interestingly, antiskyrmions have been also observed in Fe$_{1.9}$Ni$_{0.9}$Pd$_{0.2}$P which has $S_4$ symmetry \cite{Karube2021}.

While concrete simulations using atomistic spin dynamics or micromagnetics (see Sec.~\ref{t:LLG_equations}) are necessary to predict the stability of different topological magnetic objects (Fig.~\ref{f:types_of_DMI}a-c) for given material parameters and external conditions (temperature, applied field etc.), it is often useful to look at the Dzyaloshinskii-Moriya micromagnetic matrix $D_{\alpha\beta}$ (which is computationally easier to obtain than to perform dynamic simulations) to see what kind of topology can be supported by the DM interaction.

It is important to note one can combine the atomistic spin dynamics (ASD) description and micromagnetic approaches to address phenomena on different length scales in the same system within the same simulation \cite{Poluektov2016,delucia2016,Poluektov2018,Mendez2020}. For example, one can describe more accurately the interaction of magnetic textures (domain walls and skyrmions) with atomistic defects (impurities, dislocations etc.) by applying ASD in the region around the defect and micromagnetics (MM) in the rest of the system. The two regions are connected by an interface region which can be constructed in different ways to provide as seamless transition between the two regions as possible \cite{Tadmor2011,Poluektov2016,Poluektov2018}. In recent work \cite{Mendez2020}, this approach was demonstrated, for example, for a skyrmion moving near a triangular defect with locally larger anisotropy under the influence of spin-transfer torque (STT). In case of weak STT, the skyrmion is pinned at the defect in the ASD-MM simulations, while pure MM calculation results in the annihilation of the skyrmion. At larger STT, the skyrmion goes past the defect in both methods, but the MM simulations show some reduction of skyrmion size which is not seen in the more accurate ASD-MM approach. Another example in that work \cite{Mendez2020} is concerned with a linear dislocation which, similarly to the previous example, can pin a skyrmion, if STT is not strong enough. The important difference is that the skyrmion can go around the dislocation several times, depending on the magnitude of STT and damping constant, meaning that at some point it moves against the direction of STT. This unusual behavior predicted by the ASD-MM simulations cannot be reproduced by purely MM approach. In general, the ASD-MM approach combines the advantages of both methods: i) the accuracy of the ASD and ii) the larger system size that can be simulated micromagnetically.

\subsection{Numerical calculation of micromagnetic parameters}

\begin{figure}
{\centering
\includegraphics[width=0.99\textwidth]{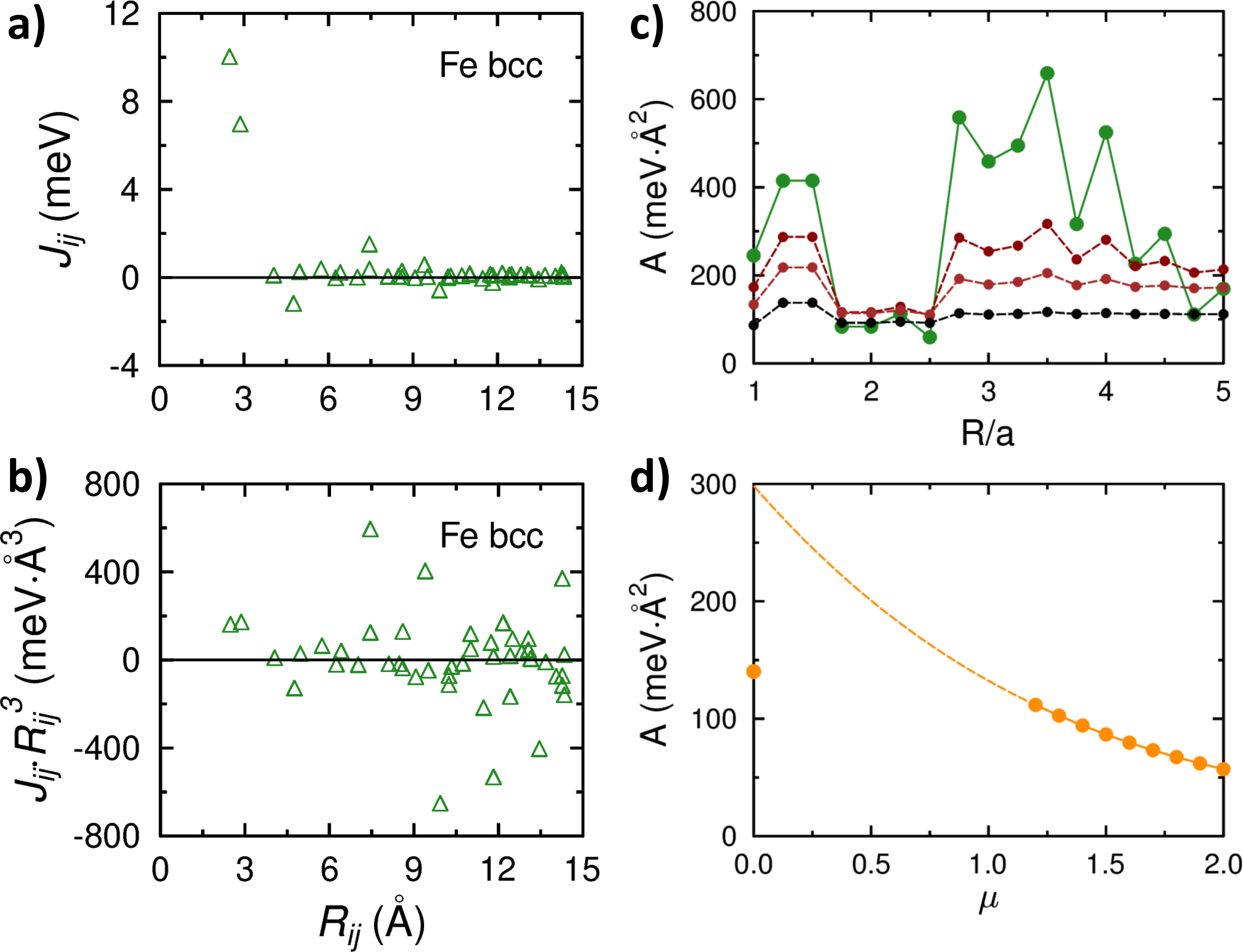}
}
\caption{a) Heisenberg interaction $J_{ij}$ in Fe \textit{bcc} for spin neighbors at different distances $R_{ij}$. b) Heisenberg interaction multiplied by $R_{ij}^3$ vs the distance $R_{ij}$; the oscillating character of the plotted quantity indicates a significant contribution of the RKKY interaction. c) Spin stiffness $A$ of Fe \textit{bcc} vs the cutoff distance for the summation in Eqn.~\ref{e:micromagnetic_parameters} with different $\mu$ values. d) Spin stiffness $A$ of Fe \textit{bcc} vs the exponential decay parameter $\mu$ (see Eqn.~\ref{e:micromagnetic_parameters}). Separate data point on the $y$-axis indicates the result of direct summation with $\mu=0$, while the dashed curve shows the extrapolation to $\mu\to 0$ limit based on the converged sums at finite $\mu$-values.}
\label{f:numerics_Fe_bcc}
\end{figure}

Despite the seeming simplicity of formulas (\ref{e:spin_stiffness}) and (\ref{e:spiralization_matrix}) defining the micromagnetic parameters $A$ and $\hat{D}$, their numerical evaluation for real systems based on the magnetic interactions calculated from first principles can be a challenge~\cite{Gayles2015,Kashin2018,Grytsiuk2019,Borisov2021,Borisov2022}. In particular, magnetic interactions in metallic systems have a long-range character. This might not be apparent at first glance, as illustrated for ferromagnetic Fe \textit{bcc} in Fig.~\ref{f:numerics_Fe_bcc}a, where it seems that the Heisenberg interaction $J_{ij}$ is insignificant for distances above $\sim\!\unit[10]{\AA}$. However, it should be kept in mind that the average number of neighbors grows with distance $R_{ij}$ and, moreover, factors of $R_{ij}^2$ and $R_{ij}$ enter the expressions for the spin stiffness and spiralization matrix (Eqns.~\ref{e:spin_stiffness} and \ref{e:spiralization_matrix}). As we discussed in Sec.~\ref{t:LKAG}, certain contributions to the Heisenberg exchange in ferromagnetic transition metals come from the $t_{2g}-t_{2g}$ orbital interactions which have an RKKY character and oscillate with distance as $\sim\!\sin(k_\mathrm{F}r)/r^3$ (Fig.~\ref{f:Jij_RKKY_in_metals}). This leads to a slow convergence of the sums (\ref{e:spin_stiffness}) and (\ref{e:spiralization_matrix}) over different neighbors with respect to the real-space cutoff distance. As shown in Fig.~\ref{f:numerics_Fe_bcc}c, the spin stiffness $A$ oscillates significantly (green curve) with varying cutoff $R$ even at larger distances of 5 lattice constants ($a$), which prevents an accurate determination of the spin stiffness parameter. In fact, the estimate of $A$ based just on the nearest-neighbor Heisenberg exchange is close to experiment ($A_\mathrm{exp} = \unit[320]{meV\cdot\AA^2}$) only by lucky chance (first green point in Fig.~\ref{f:numerics_Fe_bcc}c).

More than 20 years ago, a solution to this problem was suggested in \cite{Pajda2001} where the idea is to introduce an exponential factor in the definitions of the micromagnetic parameters:
\begin{equation}
  A = \frac{1}{2}\sum_{j\neq i} J_{ij}R^2_{ij}\,e^{-\mu R_{ij}}, \hspace{2pt} D_{\alpha\beta} = \sum_{j\neq i} D_{ij}^\alpha R_{ij}^\beta \,e^{-\mu R_{ij}}.
  \label{e:micromagnetic_parameters}
\end{equation}

Here, parameter $\mu$ is chosen to be positive, so that the exponential factors lead to faster decaying terms in the sums (\ref{e:micromagnetic_parameters}) and improve the numerical convergence with respect to the real-space cutoff. Converged results are obtained starting from a certain value $\mu_c$, which is usually around $1.0-3.0$ (depending on the system), and based on a set of $A(\mu)$ and $\hat{D}(\mu)$ calculated at different $\mu \geq \mu_c$ one can extrapolate to the limit $\mu \rightarrow 0$. The extrapolation can be done using a 3$^\mathrm{rd}$-order polynomial or exponential functions of $\mu$, as discussed in the literature, for example, for lacunar spinel GaV$_4$S$_8$ \cite{Borisov2023a} and B20 compounds~\cite{Borisov2022}.

For the simpler, yet illustrative, example considered in this review, Fe \textit{bcc}, this procedure gives $\unit[298]{meV\cdot\AA^2}$ for the spin stiffness (see the dashed line crossing the $y$-axis at $\mu=0$ in Fig.~\ref{f:numerics_Fe_bcc}d) which is quite close to the measured values of $\unit[(280-330)]{meV\cdot\AA^2}$~\cite{Pauthenet1982a,Pauthenet1982b}. In contrast, the direct summation of Eqn.~\ref{e:spin_stiffness} for spin neighbors within a distance of up to 5 lattice constants gives the estimate around $\unit[140]{meV\cdot\AA^2}$ which is more than a factor of two smaller. Another example is the B20 compound FeGe for which very different theory estimates of the spin stiffness and micromagnetic DM parameter were obtained in the literature (see discussion in \cite{Grytsiuk2019}). These micromagnetic parameters show again large oscillations as functions of the real-space cutoff distance (see Fig.~7 in \cite{Grytsiuk2019}) and the theory estimates are quite far from the measured values. The reason for this discrepancy is so far unclear, but one may speculate that higher-order magnetic interactions may play a role in this system as well as the presence of multiple sublattices  \cite{Grytsiuk2019}.

\vspace{5pt}
The method outlined above is one way of determining the micromagnetic parameters from first principles and relies on the real-space representation, as can be seen from formulas (\ref{e:spin_stiffness}) and (\ref{e:spiralization_matrix}). There are other methods as well, for example, the one discussed in~\cite{Liechtenstein1984,LKAG1987}, which is suitable for periodic crystalline systems and based on the $k$-space representation. The idea is to calculate the energy $\delta E = D_{\alpha\beta}q_\alpha q_\beta$ of a spin spiral with a small $q$-vector using the multiple-scattering theory which leads to the expression for the spin stiffness:
\begin{equation}
    D_{\alpha\beta} = \frac{1}{4\pi} \sum\limits_k \int\limits_{-\infty}^{E_\mathrm{F}} \mathrm{d}E \,\mathrm{Im} \,\mathrm{Tr}_L (t_\uparrow^{-1} - t_\downarrow^{-1})^2 \frac{\partial T_\uparrow^k}{\partial k_\alpha} \frac{\partial T_\downarrow^k}{\partial k_\beta},
    \label{e:spin_stiffness_k}
\end{equation}

where $T$ is the scattering path operator in the multiple-scattering theory \cite{Gyorffy1979}. Accurate calculation of $D_{\alpha\beta}$ depends now on the number of $k$-points used for the Brillouin zone summation in this expression, instead of the real-space cutoff in Eqns.~\ref{e:spin_stiffness} and \ref{e:spiralization_matrix}. Another advantage of this formulation is that one can describe disordered alloys within the coherent-potential approximation. Overall, this method provides estimates of the spin stiffness which are similar to the real-space method discussed above and which lie within the range of measured values. Similar approach for calculating the spin stiffness in the $k$-representation has been developed recently using the transport theory \cite{Turek2020}.

\section{Magnetization dynamics}
\label{t:LLG_equations}

\subsection{Basic equations}

Deriving the effective spin or micromagnetic models of magnetic systems is just one of the steps towards the actual modelling which is supposed to address the magnetic ground state at given conditions as well as the magnetization dynamics when these conditions are varied in time. This is the motivation of this final chapter of the review and we will consider here the fundamental aspects of magnetization dynamics and, more specifically, examples of different magnetic states that are observed in systems with Dzyaloshinskii-Moriya interaction. Theory successes in modelling such systems from first principles and the perspectives on predicting new systems with non-collinear or even topologically non-trivial magnetism will be discussed as well.

Similarly to the way that an effective spin model can be derived from the electronic problem, the equations describing the dynamics of such spin models can be derived from the time-dependent quantum-mechanical Kohn-Sham equations. The key ingredient of this derivation (discussed in detail in book \cite{Eriksson2017}) is a term proportional to $\hat{\sigma}\cdot \vec{B}_\mathrm{eff}$ which describes the spin-dependent part of the Hamiltonian and contains the effective magnetic field due to intrinsic magnetic interactions and external field. Another important aspect is the assumption that the magnetization density is well localized and locally collinear around each atom, such that its dynamics can be characterized by a single vector $\vec{m}_i$ indicating the local magnetization direction. This \textit{atomic moment approximation} is quite accurate for most systems, with a few exceptions being \textit{fcc} Pu \cite{Nordstrom1996} and Cr monolayers \cite{Sharma2007}, and allows to represent approximately the complexity of the spin density distribution in a given system by a discrete set of magnetic moments $\vec{m}_i$ on different atomic sites. Comprehensive and didactic discussion of the derivation of atomistic spin dynamics equations and many other aspects of magnetization dynamics are covered in paper~\cite{uppasd} and book~\cite{Eriksson2017}.

Both in the atomistic and in micromagnetic models, the magnetization dynamics can be described by the Landau-Lifshitz-Gilbert (LLG) equation \cite{Landau1935,Gilbert2004}:
\begin{equation}
    \frac{\partial \vec{m}_i}{\partial t} = -\frac{\gamma}{1 + \alpha^2} \left[ \vec{m}_i \times \vec{B}_i + \frac{\alpha}{m}\,\vec{m}_i \times (\vec{m}_i \times \vec{B}_i) \right]
    \label{e:LLG_equation}
\end{equation}

which describes the time-evolution of magnetic moments $\vec{m}_i(t)$ on different atoms, molecules or other effective magnetic entities (numbered with index ``$i$'') or in different micromagnetic regions. The system-specific parameters here are the Gilbert damping constant $\alpha$, which characterizes the energy dissipation, and the effective field $\vec{B}_i$. The later contains contributions from the external magnetic field as well as the Heisenberg and Dzyaloshinskii-Moriya interactions, dipole-dipole energy, on-site anisotropy etc. At finite temperature $T$, random field $\vec{B}_{i,fl}$ proportional to $\sqrt{\alpha T}$ is added to the effective field $\vec{B}_i$ to include the fluctuation effects in the simulation. The connection between the damping and fluctuations in the spin system characterized by the quantities $\alpha$ and $\vec{B}_{i,fl}$ is not accidental and is a consequence of the fundamental \textit{fluctuation-dissipation theorem} \cite{Callen1951,Kubo1966}. The resulting stochastic differential equations are solved using the method of Langevin dynamics \cite{vamKampen2007} by rewriting the LLG equation with the fluctuating field as a Fokker-Planck equation (solution methods for these equations can be found, for example, in \cite{Risken1989}).

It is the accuracy of calculating the effective field $\vec{B}_\mathrm{eff}$, which mostly determines the predictive power of the atomistic spin dynamics simulation for a given system. The $\vec{B}_\mathrm{eff}$ field is determined from the spin Hamiltonian as $-\frac{\partial H}{\partial \vec{m}_i}$ \cite{Landau1935}. Correspondingly, one would get different expressions for the atomistic and micromagnetic cases, where the starting points are the Heisenberg model (\ref{e:Heisenberg_model}) and micromagnetic functional (\ref{e:micromagnetic_functional}). Different contributions to the effective field are summarized in Table~\ref{tab:effective_field} for these two approaches. Higher-order spin-spin interactions highlighted in Sec.~\ref{t:higher_order_spin} would make further contributions to the effective field and can be evaluated using the general gradient formula mentioned above. It should be noted that, in the approximation of constant-moment length ($|\vec{m}_i| = 1$), the component of the effective field parallel to the magnetic moment $\vec{m}_i$ does not exert any torque and vanishes from the LLG equation (\ref{e:LLG_equation}). Another important aspect of first-principles spin dynamics is concerned with the concept of the \textit{constraining field}, which, as we already mentioned in Sec.~\ref{t:LKAG}, is used to stabilize a non-collinear spin configuration different from the magnetic ground state. In a recent work~\cite{Streib2020}, the main conclusion is that the constraining field is not always equal to the gradient of the Hamiltonian, if the later contains mean-field-like contributions or is based on density functional theory. This can also have important consequences when the magnetic exchange parameters are determined from the constraining field approach~\cite{Streib2020,Streib2021,Streib2022}.

\setlength{\tabcolsep}{7pt}
\renewcommand{\arraystretch}{1.5}
\begin{table}
\caption{\label{jlab1}Summary of the effective field related to the Heisenberg, Dzyaloshinskii-Moriya (DM), dipole-dipole interactions and uniaxial anisotropy for the atomistic and micromagnetic pictures.}
\footnotesize
\begin{tabular}{@{}lll}
\br
Origin&Atomistic picture&Micromagnetic picture\\
\mr
Heisenberg exchange & $+\sum\limits_{j\neq i} J_{ij} \vec{S}_j$ & $-A\nabla^2 \vec{m}$ \\[5pt]
DM interaction & $-\sum\limits_{j\neq i} \vec{D}_{ij}\times \vec{S}_j$ & $-( \hat{D}\,\vec{\nabla} ) \times \vec{m}$ \\[5pt]
Dipole-dipole energy & $\frac{\mu_0}{2\pi}\sum\limits_{j\neq i} \frac{1}{r_{ij}^3} \left( 3\,\hat{r}_{ij} (\vec{m}_j\cdot \hat{r}_{ij}) - \vec{m}_j \right) $ & $-\frac{\mu_0}{4\pi} \int \vec{\nabla}\vec{\nabla}' \frac{1}{|\vec{r}-\vec{r}'|} \vec{M}(\vec{r}\,')\,\mathrm{d}\vec{r}\,'$ \\[5pt]
Uniaxial anisotropy & $-K_U (\vec{S}_i\cdot\vec{e})\,\vec{e}$ & $-K_U (\vec{m}\cdot\vec{e})\,\vec{e}$ \\
\br
\end{tabular}\\
\label{tab:effective_field}
\end{table}

\subsection{Prediction of magnetic properties and textures}
\label{t:magn_predictions}

\begin{figure}
{\centering
\includegraphics[width=0.99\textwidth]{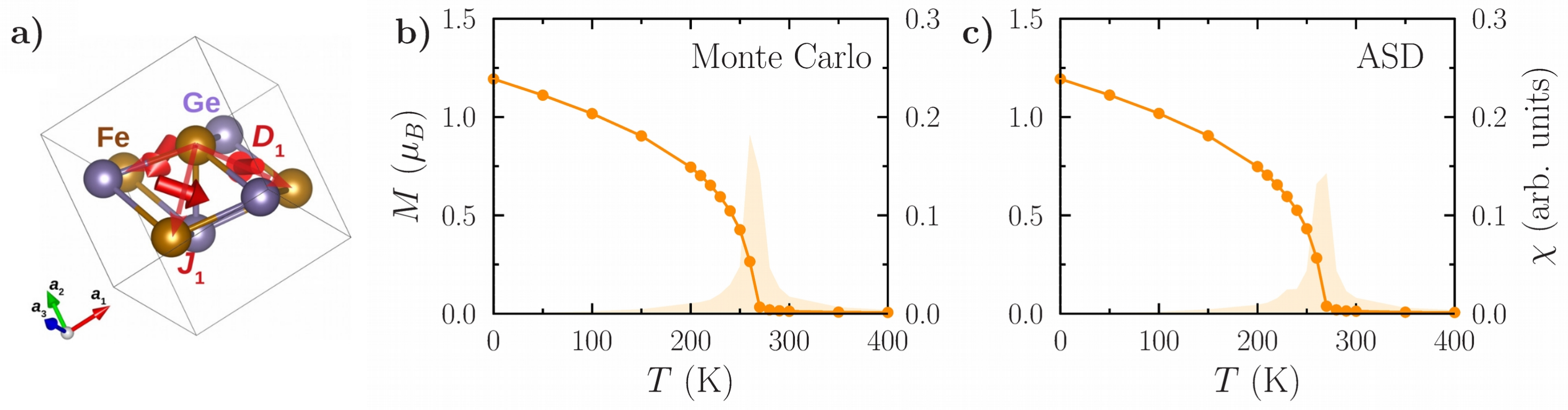}
\includegraphics[width=0.99\textwidth]{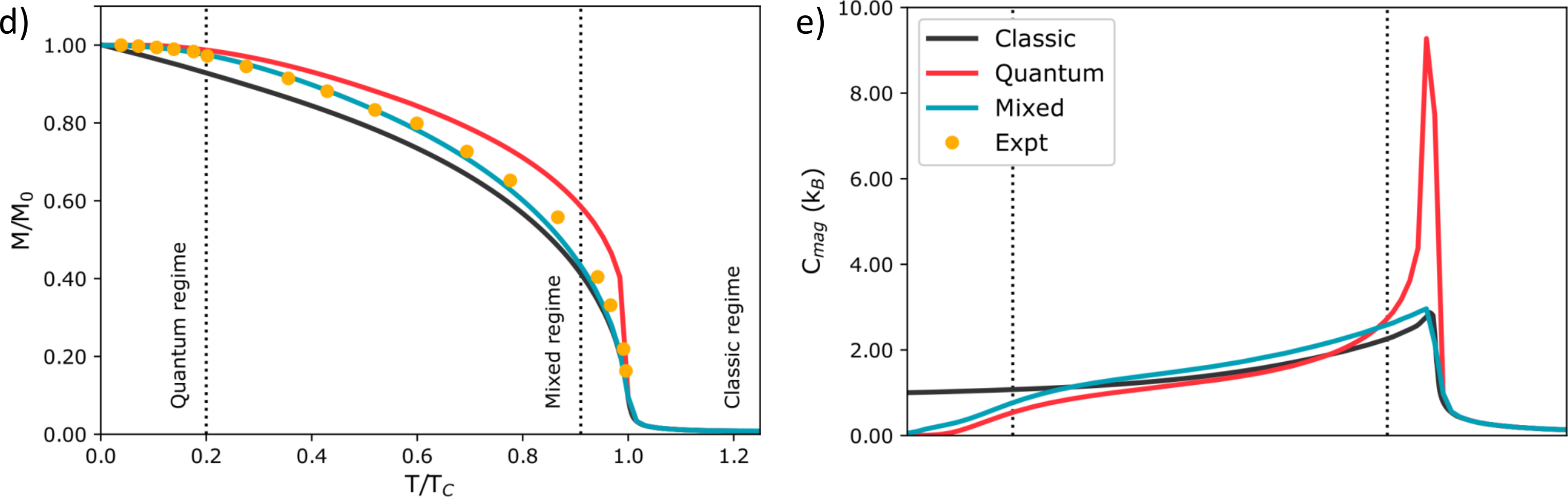}
}
\caption{a) Crystal structure of the skyrmionic B20 compound FeGe and the DM vectors for the nearest-neighbor bonds (reproduced from \cite{Borisov2021}). Temperature-dependent magnetization and susceptibility are obtained from the b) Monte Carlo and c) atomistic spin dynamics (ASD) simulations using the spin model from \cite{Borisov2021}. d) Magnetization and e) heat capacity of ferromagnetic \textit{hcp} Gd as functions of temperature obtained from Monte Carlo simulations based on different types of statistics (classical, quantum and mixed); plot a) is reproduced from \cite{Borisov2021} and plots d) and e) are reproduced from \cite{Vieira2022}; experimental data points are from Ref.~\cite{Nigh1963}.}
\label{f:FeGe_M_vs_T}
\end{figure}

Based on the effective spin model, one can calculate the corresponding thermodynamic quantities using either the Monte Carlo (MC) stochastic sampling or the atomistic spin dynamics (ASD) described by the LLG equation. In case of MC approach, a series of quasi-random perturbations of the spin configuration is generated using the Metropolis algorithm (more details in \cite{Janke1996}, pages 17--25) and the thermodynamic properties are calculated as averages over this data set, which is supposed to represent the variety of system configurations and their probabilities to a good accuracy. The later depends on the number of Monte Carlo samples. In Fig.~\ref{f:FeGe_M_vs_T}b, the MC prediction for the temperature-dependent magnetization $M(T)$ and susceptibility $\chi(T)$ is shown for bulk FeGe compound (structure in Fig.~\ref{f:FeGe_M_vs_T}a), which is one of the skyrmionic compounds \cite{Lundgren1968,Wappling1968,Lebech1989,Yu2011,Siegfried2017,Zheng2018}. The simulation is based on the spin model obtained in \cite{Borisov2021}, which includes the Heisenberg and DM exchange interactions between spin neighbors up to a distance of 3 lattice parameters ($\sim\!\unit[14.1]{\AA}$). The magnetization $M(T)$ decreases monotonically with temperature and approaches zero around $T_\mathrm{C}\sim\!\unit[260]{K}$, while the magnetic susceptibility reaches the maximal value at that temperature, which is a typical behavior of ferromagnets. Similar result is obtained in ASD simulations with the same exchange interaction parameters (Fig.~\ref{f:FeGe_M_vs_T}c). The theoretical estimate for the critical temperature is close to the experimental value $T_\mathrm{C}\sim\!\unit[280]{K}$ \cite{Lundgren1968,Wappling1968,Lebech1989,Siegfried2017}.

It is important to note that in the classical spin simulations the decrease of magnetization at low temperatures is linear, which is in disagreement with the quantum picture where magnon excitations reduce $M(T)$ by a quantity $\sim\!T^{3/2}$ in case of 3D systems. This artifact of classical simulations also leads to a constant finite heat capacity at low temperatures, which reflects the classical equipartition theorem. However, in the quantum picture the heat capacity should approach zero as $T^{3/2}$. Ways of correcting this inconsistency were suggested, for example, in \cite{Woo2015,Evans2015,Bergqvist2018} where one of the ideas is to rescale the simulation temperature based on the average magnon energy.

The ASD simulation with this pure quantum statistic predicts well only the low-temperature part of the $M(T)$ curve \cite{Woo2015,Evans2015,Bergqvist2018}, while the classical statistics is reasonably accurate at higher temperatures, around and above the critical temperature $T_\mathrm{C}$. In a recent work \cite{Vieira2022}, the authors proposed a combination of both approaches using a mixed statistics which interpolates smoothly between the quantum and classical limits. That is, in the Monte Carlo simulations at low temperature the probability $W_\mathrm{qt}(\Delta E, T)$ of changing the state of the system, accompanied by energy change $\Delta E$, is defined using quantum statistics \cite{Woo2015,Evans2015,Bergqvist2018}, while at high temperatures near and above the magnetic transition the probability $W_\mathrm{cl}(\Delta E, T)$ is described by the Boltzmann distribution. In-between the two limits $T=0$ and $T\sim T_\mathrm{C}$, the probability is a linear interpolation with a factor $\alpha = 1 - \frac{T}{T_\mathrm{C}}$: $W_\mathrm{mix}(\Delta E, T) = \alpha W_\mathrm{qt}(\Delta E, T) + (1 - \alpha) W_\mathrm{cl}(\Delta E, T)$. As a result, the mixed-statistics Monte Carlo simulation predicts a temperature dependence of the magnetization $M(T)$, on the example of ferromagnetic \textit{hcp} Gd, which is in a much better agreement with measurements (Fig.~\ref{f:FeGe_M_vs_T}d) and a correct behavior of the heat capacity (Fig.~\ref{f:FeGe_M_vs_T}e).

\begin{figure}
{\centering
\includegraphics[width=0.99\textwidth]{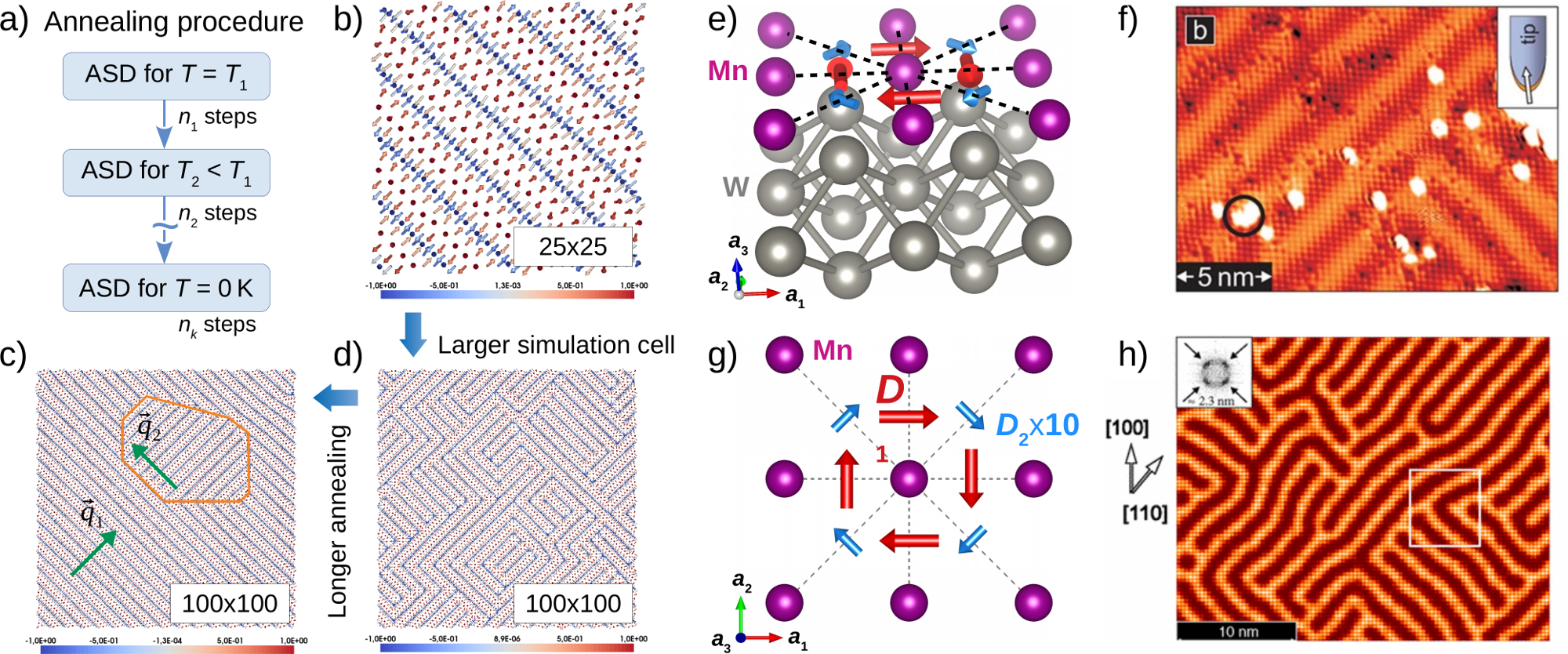}
}
\caption{a) Schematic illustration of the annealing procedure in atomistic spin dynamics (ASD) simulations. Relaxation of the spin structure is done at gradually lowered temperature until $T=\unit[0]{K}$ is reached. b) Resulting spin structure (single-domain spin spiral) for Mn/W(001) layered 2D system with uniaxial anisotropy simulated using a $(25\times25)$ cell. d) Spin configuration obtained in a larger simulation cell reveals coexisting spin-spiral domains oriented in [110] and [1-10] directions. c) Allowing for longer simulation times and more temperature steps in annealing leads to larger spin-spiral domains. e) Crystal structure of Mn/W(001) layered system with the DM vectors shown for the nearest- and next-nearest-neighbor bonds. f) Experimentally measured magnetic state \cite{Ferriani2018} showing slightly distorted spin-spiral domains. g) Top view of the Mn monolayer structure and DM vectors. h) Magnetic structure predicted by Monte Carlo simulations at $\unit[13]{K}$. Figures e) and g) are reproduced from \cite{Borisov2021} and figures f) and h) are reproduced from \cite{Ferriani2018}.}
\label{f:MnW001}
\end{figure}

Let us discuss how the atomistic spin dynamics (ASD) and micromagnetic approach can be used to search for the magnetic ground state of a given system. In both formalisms, a possible strategy is to perform the \textit{annealing simulation} which starts from a random magnetic configuration at high temperature and the magnetic dynamics is simulated using the LLG equation (\ref{e:LLG_equation}) while the temperature is gradually lowered down to zero (illustration in Fig.~\ref{f:MnW001}a). Such simulations usually result in a local energy minimum which can bare many features of the actual magnetic ground state. We demonstrate this procedure on the concrete example of Mn monolayer on W(001) surface. Literature studies \cite{Udvardi2008,Freimuth2014,Ferriani2018} indicate a particularly strong DM interaction in this layered system induced by the Mn/W(001) interface (similarly to Mn/W(110) system \cite{Bode2007,Udvardi2008}) with large spin-orbit coupling due to W and inversion symmetry breaking. For the nearest-neighboring spins, the DMI is almost 50\% of the Heisenberg exchange. Also, the uniaxial on-site anisotropy is very large $\sim\!\unit[2.5]{meV}$ per Mn site, according to \textit{ab initio} estimates from \cite{Borisov2021}. Using the first-principles values of the magnetic interactions and other parameters from that work \cite{Borisov2021}, we compare further below the resulting ground state found in ASD. For the smallest considered simulation cell of dimensions $(25\times 25)$ the ASD simulation predicts a single-domain spin-spiral phase (Fig.~\ref{f:MnW001}b) with a small spatial period, similarly to experimental result \cite{Ferriani2018}. When a larger $(100\times 100)$ simulation cell is used, the spin structure splits into several domains oriented either in [110] or [1-10] directions. Both domain types have the same energy and, for that reason, can coexist in the system. The distribution of domains depends on the initial spin configuration, which is chosen randomly, and the cooling rate. Importantly, for lower cooling rate, i.e. longer simulation time at each temperature, the number of domains decreases (example with just two domains in shown in Fig.~\ref{f:MnW001}c) and the required number of time steps in the simulation depends on the cell size. For that reason, single-domain state is achieved faster in smaller $(25\times 25)$ cell, while the larger $(100\times 100)$ cell splits into a multi-domain state for the same cooling rate. For comparison, similar spin configuration is predicted by Monte Carlo (MC) simulations at $\unit[13]{K}$ in \cite{Ferriani2018} shown in Fig.~\ref{f:MnW001}h in comparison with the spin-polarized scanning transmission microscopy image (Fig.~\ref{f:MnW001}f) at the same temperature. The magnetic state predicted by the MC simulations consists again of spin-spiral domains oriented in [110] and [1-10] directions, in accordance with measurements.

Apart from spin spirals, the DM interaction can also stabilize topologically non-trivial magnetic states, for example, skyrmionic phases with either isolated skyrmions or skyrmion lattices. Being originally proposed as topological defects in high-energy physics by Tony Skyrme in 1961 \cite{Skyrme1961}, skyrmions have been identified also in other research fields, e.g.~liquid crystals \cite{Wright1989} and superconductors \cite{Abrikosov2004}. Skyrmions are compact magnetic objects formed by atomic spins which are winding around a certain direction, such that the spins in the center of the skyrmion are opposite to those at its rim and the intermediate spins rotate smoothly in-between the center and the rim (Fig.~\ref{f:types_of_DMI}a-c). This winding is associated with a non-trivial topology which can be characterized by the topological charge:
\begin{equation}
    Q = \frac{1}{4\pi} \int \vec{m} \cdot \left( \frac{\partial \vec{m}}{\partial x} \times \frac{\partial \vec{m}}{\partial y} \right) \mathrm{d}x\,\mathrm{d}y,
    \label{e:topological_number}
\end{equation}

where vector $\vec{m}=\vec{m}(\vec{r})$ describes the direction of magnetization at a given point $\vec{r}$ in space. For Bloch and N\'eel skyrmions (Fig.~\ref{f:types_of_DMI}a,b) the topological charge equals $+1$, while it is $-1$ for antiskyrmions (Fig.~\ref{f:types_of_DMI}c). Several material classes with different types of topological magnetic textures are known (see e.g.~reviews \cite{Kanazawa2017,Bihlmayer2018,EverschorSitte2018,Goebel2021,Zhang2023}) and searching for new systems of that kind is an on-going effort in the research community.

Theoretical simulations based on first-principles approaches can provide useful insights into the physical mechanism of stability of topological magnetism and even make quantitative predictions of materials properties. In the following, we consider a few literature examples illustrating this statement.

The first example is the nanoskyrmion lattice observed in Fe monolayer on Ir(111) surface. Spin-polarized scanning transmission tunneling microscopy experiments showed that the Fe moments in this system form a square lattice of skyrmions with a period of $\sim\!\unit[1]{nm}$ \cite{Heinze2011}. Interestingly, this skyrmion lattice is incommensurate with the underlying hexagonal lattice of Fe sites. From the analysis of experimental data, one can conclude that the observed magnetic state is a superposition of two spin spirals with $Q$-vectors $\vec{Q}_1$ and $\vec{Q}_2$ which are at $92.2^\mathrm{o}$ to each other and have the same magnitude $0.277\times 2\pi/a$. Theoretical analysis in \cite{Heinze2011} based on an effective spin model (\ref{e:spin12_model}) with an addition of DM interaction revealed that the nearest-neighbor four-spin interaction favors the double-$Q$ magnetic state combining the $\vec{Q}_1$ and $\vec{Q}_2$ spirals. This magnetic superposition can lead to either skyrmions or antiskyrmions, which in the non-relativistic case are degenerate in energy, but due to the specific symmetry of the DM interaction in this system skyrmions are lower in energy. Using the magnetic interaction parameters calculated from first principles, the authors quantified the effect of the Heisenberg, four-spin and DM interactions on the formation of the nanoskyrmion lattice and obtained a good agreement with experiment in terms of the magnitude of the $\vec{Q}_1$ and $\vec{Q}_2$ vectors that minimize the magnetic energy.
\begin{figure}
{\centering
\includegraphics[width=0.99\textwidth]{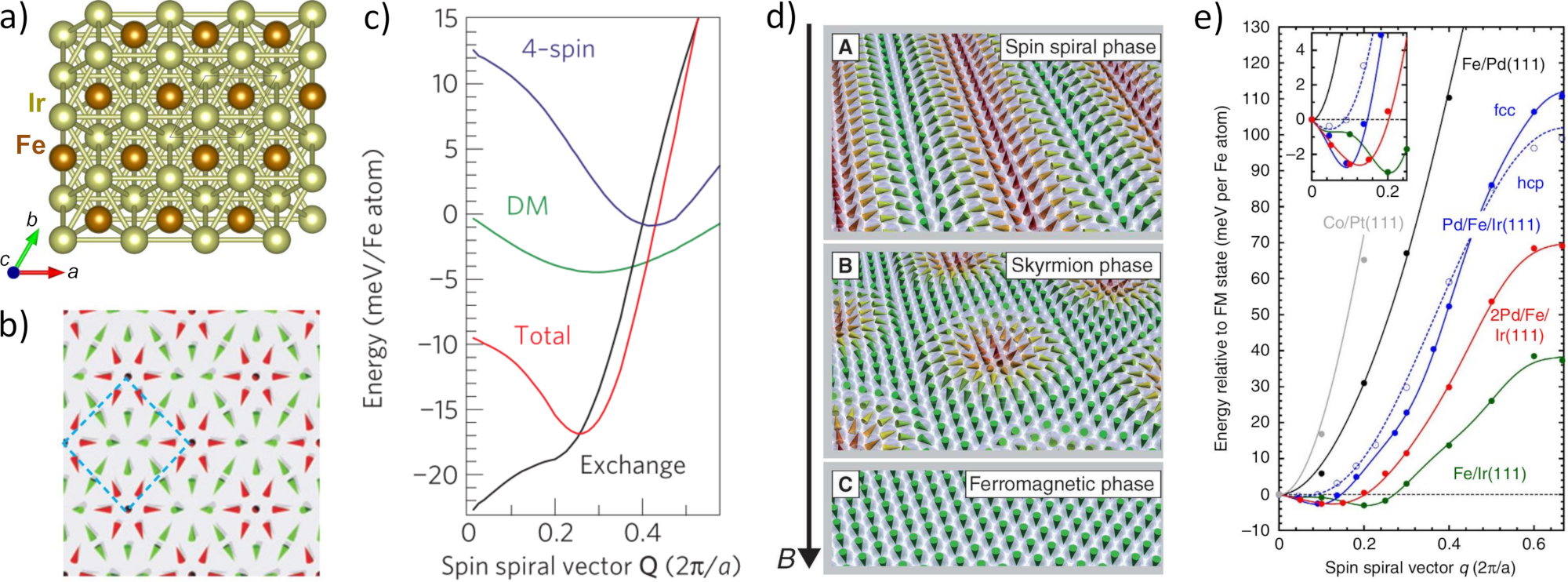}
}
\caption{a) Crystal structure of Fe monolayer on Ir(111) surface. b) Schematic illustration of the nanoskyrmion lattice found in this system. c) Magnetic energy calculated for the double-$Q$ state as a function of the $Q$-vector magnitude; contributions of the Heisenberg exchange, four-spin and DM interactions to the total energy are shown. d) Schematic representation of the magnetic phases observed in Pd/Fe/Ir(111) multilayer at increasing applied magnetic field. e) Calculated spin-spiral energy for various multilayers as functions of the $Q$-vector. Figures b-c) are reproduced from \cite{Heinze2011} and figure d)~-- from \cite{Romming2013} and e)~-- from \cite{Dupe2014}.}
\label{f:Pd_Fe_Ir111}
\end{figure}

It is interesting to see how one can drastically modify the ground state of this layered nanoskyrmion system by simply placing a single Pd layer on top of Fe. Experimentally \cite{Romming2013}, it is known that this changes the ground state to a spin spiral with a period around $\unit[6-7]{nm}$ which can be transformed to a skyrmion lattice of similar length scale in applied magnetic field around $\unit[1]{T}$ (Fig.~\ref{f:Pd_Fe_Ir111}d). Theoretical calculations in \cite{Dupe2014} revealed that the Fe/Ir(111) interface still dominates the DM interaction, which determines the chirality of spin spirals, but the presence of the Pd layer enhances the ferromagnetic Heisenberg exchange between the nearest-neighbor Fe spins by more than a factor of 2, compared to the Fe/Ir(111) system. This is based on spin-spiral energies calculated for both systems (Fig.~\ref{f:Pd_Fe_Ir111}e). Furthermore, there are competing antiferromagnetic interactions for more distant neighbors which favor a spin spiral state even without the spin-orbit coupling. At zero field, both factors lead to stability of spin-spiral state instead of skyrmions. Another statement from that work is that the agreement with experiment in terms of spiral wavelength, skyrmion size and critical magnetic fields for these two phases is especially good for the \textit{hcp} stacking of the Pd layer on top of Fe layer. On the other hand, the \textit{fcc} stacking would lead to deeper energy minimum of a spin spiral (Fig.~\ref{f:Pd_Fe_Ir111}e) leading to higher magnetic field required to stabilize a skyrmion lattice. It is worth noting that the strong spin-orbit coupling (SOC) due to the Ir(111) surface is essential for the skyrmion stability, as demonstrated by calculations for Fe/Pd(111) multilayer where SOC is an order of magnitude weaker and is not enough to stabilize even spin spirals (Fig.~\ref{f:Pd_Fe_Ir111}e).

These literature studies of Fe/Ir(111) and Pd/Fe/Ir(111) systems demonstrate that the multiscale approach can be useful for understanding the physical mechanism of the observed magnetic phenomena and how they react to variations of the material properties. With some adjustment of theory approximations and models, the agreement between theory and experiment can be achieved even on the quantitative level.

\begin{figure}
{\centering
\includegraphics[width=0.99\textwidth]{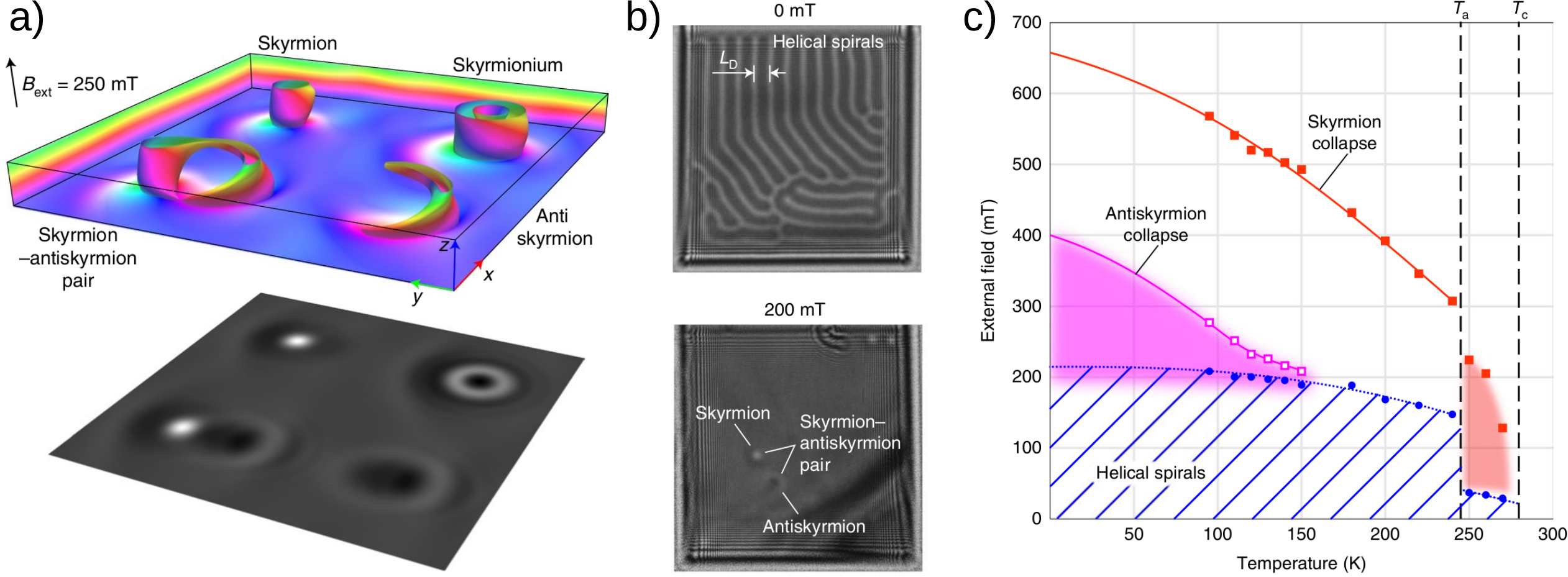}
}
\caption{a) Isosurfaces ($m_z = 0$) of the spin textures in FeGe system with dimensions $\unit[(512\times512\times50)]{nm^3}$ from micromagnetic simulations showing skyrmion, antiskyrmion, skyrmion–antiskyrmion pair and skyrmionium. The figure below is the calculated over-focused Lorentz TEM image for these spin textures. b) Measured Lorentz TEM images of FeGe at $\unit[95]{K}$ in zero field and $H=\unit[200]{mT}$ (sample size: $\unit[1]{\mu m}\times \unit[1]{\mu m}$). c) Combined experimental and theoretical phase diagram showing the skyrmionic, antiskyrmionic and helical spiral phases of FeGe; the symbols are the experimental data points; the red-shaded region represents the skyrmion lattice phase and the vertical dashed line at $\unit[287]{K}$ shows the Curie temperature. All figures are reproduced from \cite{Zheng2022}.}
\label{f:B20s_micro}
\end{figure}

Let us now consider another example where theory predicts an interesting magnetic phenomenon which is later on confirmed experimentally. Theory analysis of 2D isotropic magnets with DMI in \cite{Kuchkin2020,BartonSinger2020} showed that not only skyrmions but also antiskyrmions can be stabilized, but in an extremely narrow range of applied magnetic field. In a more recent work \cite{Zheng2022}, micromagnetic simulations revealed that in the 3D generalization of this model antiskyrmions become more stable due to the dipolar field contribution leading to easy-plane anisotropy and the possibility of non-collinearity in the direction perpendicular to the skyrmion plane, which enhances the DMI energy. However, the skyrmion-antiskyrmion pairs are less stable than the isolated topological magnetic particles. Furthermore, skyrmionium has been found to be a stable solution in the 3D simulations (Fig.~\ref{f:B20s_micro}a). These theory predictions were confirmed successfully by experiments in the same work \cite{Zheng2022} done on thin FeGe samples with a thickness of $\sim\!\unit[70]{nm}$. Lorentz transmission electron microscopy (TEM) images (Fig.~\ref{f:B20s_micro}b) show clearly how the original zero-field helical spiral state transforms into skyrmions and antiskyrmions (Fig.~\ref{f:B20s_micro}c) in applied magnetic field (skyrmionium is also stabilized at a slightly stronger field) and how the pairs of these topological particles annihilate when in close proximity to each other (see futher figures in \cite{Zheng2022}). The simulated TEM images (Fig.~\ref{f:B20s_micro}a) are in a good agreement with the measurements, which helps to identify the nature of topological magnetic textures observed in experiment. It turns out that size effects related to the thickness of the FeGe sample are important for the formation of antiskyrmions, which can be stabilized if the thickness does not exceed the spiral wavelength. Thicker samples, on the other hand, only show usual skyrmions \cite{Zheng2022}.

\subsection{Spin-lattice dynamics}

It is worth noting that the LLG equations (\ref{e:LLG_equation}) are derived in the adiabatic limit where the spin degrees of freedom are assumed to be much faster than the ionic degrees of freedom, which is true in many cases. Material-specific spin dynamics in this approximation with parameters obtained from first principles has been considered in \cite{Niu1999,Brown2001,Qian2002,Nowak2005,uppasd,Bhattacharjee2012a,Bhattacharjee2012b}, while some of the first works \cite{Antropov1995,Antropov1996a,Antropov1996b} on spin dynamics included also the coupling between the spin and ionic sub-systems.

More recently, methodology for simulating coupled spin-lattice phenomena is suggested in \cite{Hellsvik2019} where the idea is to combine the LLG equations (\ref{e:LLG_equation}) and molecular dynamics simulation in the harmonic-phonon approximation. The full Hamiltonian of the system, discussed in that work \cite{Hellsvik2019}, consists then of the pure spin ($H_{S}$) and lattice ($H_{L}$) parts and the coupling term $H_{LS}$. The spin part is based on the original Heisenberg model where the exchange tensor $J_{ij}^{\alpha\beta}$ is expanded in a Taylor series as a function of small displacements $\vec{u}$ of different atoms:
\begin{equation}
    H_{spin} = -\frac{1}{2}\sum\limits_{ij} J_{ij}^{\alpha\beta}m_i^\alpha m_j^\beta - \frac{1}{2}\sum\limits_{ijk} \frac{\partial J_{ij}^{\alpha\beta}}{\partial u_k^\mu}u_k^\mu m_i^\alpha m_j^\beta - \frac{1}{4}\sum\limits_{ijkl} \frac{\partial^2 J_{ij}^{\alpha\beta}}{\partial u_k^\mu \partial u_l^\nu} u_k^\mu u_l^\nu m_i^\alpha m_j^\beta
    \label{e:spin_Hamiltonian_HSS}
\end{equation}

This expression contains the pure spin part $H_S$ and a part of the spin-lattice coupling in the form of cross-terms depending both on the magnetic moments and ionic displacements. The lattice contribution to energy contains the kinetic energy of the nuclei and takes into account the dependence of the ionic forces on the magnetic configuration by expanding the force constants into a Taylor series with respect to small deviations of magnetic moments $\vec{m}_i$ from the equilibrium state:
\begin{eqnarray}
    H_{lattice} & = \frac{1}{2}\sum\limits_k M_k \upsilon_k^\mu \upsilon_k^\mu + \frac{1}{2} \sum\limits_{kl} \Phi_{kl}^{\mu\nu} u_k^\mu u_l^\nu + \frac{1}{2} \sum\limits_{ikl} \frac{\partial \Phi_{kl}^{\mu\nu}}{\partial m_i^\alpha} m_i^\alpha u_k^\mu u_l^\nu + \\[5pt]
     & + \frac{1}{4}\sum\limits_{ijkl} \frac{\partial^2 \Phi_{kl}^{\mu\nu}}{\partial m_i^\alpha m_j^\beta} m_i^\alpha m_j^\beta u_k^\mu u_l^\nu
     \label{e:lattice_Hamiltonian_HLL}
\end{eqnarray}

Combining both expansions (\ref{e:spin_Hamiltonian_HSS}) and (\ref{e:lattice_Hamiltonian_HLL}) allows to write the spin-lattice coupling in the following general way:
\begin{equation}
    H_{LS} = \frac{1}{2} \sum\limits_{ikl} \Theta_{ikl}^{\alpha\mu\nu} m_i^\alpha u_k^\mu u_l^\nu - \frac{1}{2} \sum\limits_{ijk} \Gamma_{ijk}^{\alpha\beta\mu} u_k^\mu m_i^\alpha m_j^\beta - \frac{1}{4}\sum\limits_{ijkl} \Lambda_{ijkl}^{\alpha\beta\mu\nu} m_i^\alpha m_j^\beta u_k^\mu u_l^\nu,
    \label{e:spin_lattice_coupling}
\end{equation}

while the pure spin and lattice parts are defined in this way:
\begin{equation}
    H_S = -\frac{1}{2}\sum\limits_{ij} J_{ij}^{\alpha\beta}m_i^\alpha m_j^\beta
    \label{e:pure_spin_Hamiltonian}
\end{equation}
\begin{equation}
    H_L = \frac{1}{2}\sum\limits_k M_k \upsilon_k^\mu \upsilon_k^\mu + \frac{1}{2} \sum\limits_{kl} \Phi_{kl}^{\mu\nu} u_k^\mu u_l^\nu
    \label{e:pure_lattice_Hamiltonian}
\end{equation}

Also, the procedure for determining all the coupling coefficients in this expression is proposed in \cite{Hellsvik2019}. The idea is to consider a supercell where one of the atoms is displaced and the magnetic interactions are calculated as functions of the displacement, making it possible to calculate the first and second derivatives in Eqn.~(\ref{e:spin_lattice_coupling}). Alternative ways to calculate these coefficients, for example, using a model for the exchange parameters were proposed in \cite{Ma2008,Ma2012,Perera2017,Assmann2019}. Evaluation method for the spin-lattice coupling coefficients using the Green function method, similar to the LKAG approach \cite{LKAG1987}, and embedded cluster method have been proposed recently in \cite{Mankovsky2022} and \cite{Lange2023}, respectively, and tested for several representative systems. For example, analysis of spin-lattice coefficients in \cite{Mankovsky2022} indicated the importance of Dzyaloshinskii-Moriya interactions in Fe \textit{bcc}, induced by lattice vibrations, for the transfer of angular momentum between the spin and lattice subsystems.

Modelling the coupled spin-lattice dynamics at finite temperature requires solving the equation of motion both for the spins and for the lattice sites simultaneously, leading to an extended system of equations compared to Eqn.~(\ref{e:LLG_equation}):
\begin{eqnarray}
    \frac{\partial \vec{m}_i}{\partial t} &= -\frac{\gamma}{1 + \alpha^2} \left[ \vec{m}_i \times (\vec{B}_i + \vec{B}_i^{fl}) + \frac{\alpha}{m}\,\vec{m}_i \times (\vec{m}_i \times (\vec{B}_i + \vec{B}_i^{fl})) \right]\\[5pt]
    \frac{\mathrm{d}\vec{v}_k}{\mathrm{d}t} &= \frac{\vec{F}_k + \vec{F}_k^{fl}}{M_k} - \nu \vec{v}_k, \hspace{10pt} \frac{\mathrm{d}\vec{u}_k}{\mathrm{d}t} = \vec{v}_k
    \label{e:LLG_and_ionic_equations}
\end{eqnarray}

The fields acting on the spin and lattice degrees of freedom are calculated in a similar way based on Hamiltonian $H_\mathrm{total} = H_S + H_L + H_{LS}$ defined by Eqns.~\ref{e:spin_lattice_coupling}--\ref{e:pure_lattice_Hamiltonian}:
\begin{equation}
    \vec{B}_i = -\frac{\partial H_\mathrm{total}}{\partial \vec{m}_i}, \hspace{10pt} \vec{F}_k = -\frac{\partial H_\mathrm{total}}{\partial \vec{u}_k}
    \label{e:effective_fields}
\end{equation}

Additional contribution to these fields comes from fluctuations at finite temperature which can be approximately described by white (Gaussian) noises with correlation functions ($\mu,\nu=x,y,z$):
\begin{eqnarray}
    \left\langle B_{i,\mu}^{fl}(t) B_{j,\nu}^{fl}(t') \right\rangle &= 2D_M \delta_{ij}\delta_{\mu\nu}\delta(t-t')\\[5pt]
    \left\langle F_{k,\mu}^{fl}(t) F_{l,\nu}^{fl}(t') \right\rangle &= 2D_L \delta_{kl}\delta_{\mu\nu}\delta(t-t')
\end{eqnarray}

The important parameters here are $D_M$ and $D_L$ which characterize the strength of fluctuations but, on the other hand, through the fluctuation-dissipation theorem \cite{Callen1951,Kubo1966} are intimately related to the energy dissipation in the spin and lattice subsystems:
\begin{equation}
    D_M = \frac{\alpha k_\mathrm{B}T}{\gamma m}, \hspace{10pt} D_L = \nu M k_\mathrm{B} T,
    \label{e:dissipation_spin_and_lattice}
\end{equation}

where $\alpha$ and $\nu$ are the spin and lattice damping parameters.

It should be noted that the possibility of continuum formulation of the equations of coupled spin-lattice dynamics has been discussed in recent work \cite{Weissenhofer2023}. The idea is similar to the micromagnetic approach (Sec.~\ref{t:micromagnetics} of this review) which introduces a continuous vector field $\vec{m}(\vec{r})$ describing the magnetization distribution. For the spin-lattice case, one can also define the mechanical energy in terms of the strain and stress tensors $\varepsilon_{\alpha\beta}(\vec{r})$ and $\sigma_{\alpha\beta}(\vec{r})$, using the well-established elasticity theory \cite{Lifshitz1986}, as well as cross-terms describing the coupling between the strain tensor and magnetization and their gradients. These cross-terms can be obtained systematically starting from the atomistic model (Eqns.~\ref{e:spin_lattice_coupling}--\ref{e:pure_lattice_Hamiltonian}) and expanding the continuous variables in real space.

\begin{figure}
{\centering
\includegraphics[width=0.99\textwidth]{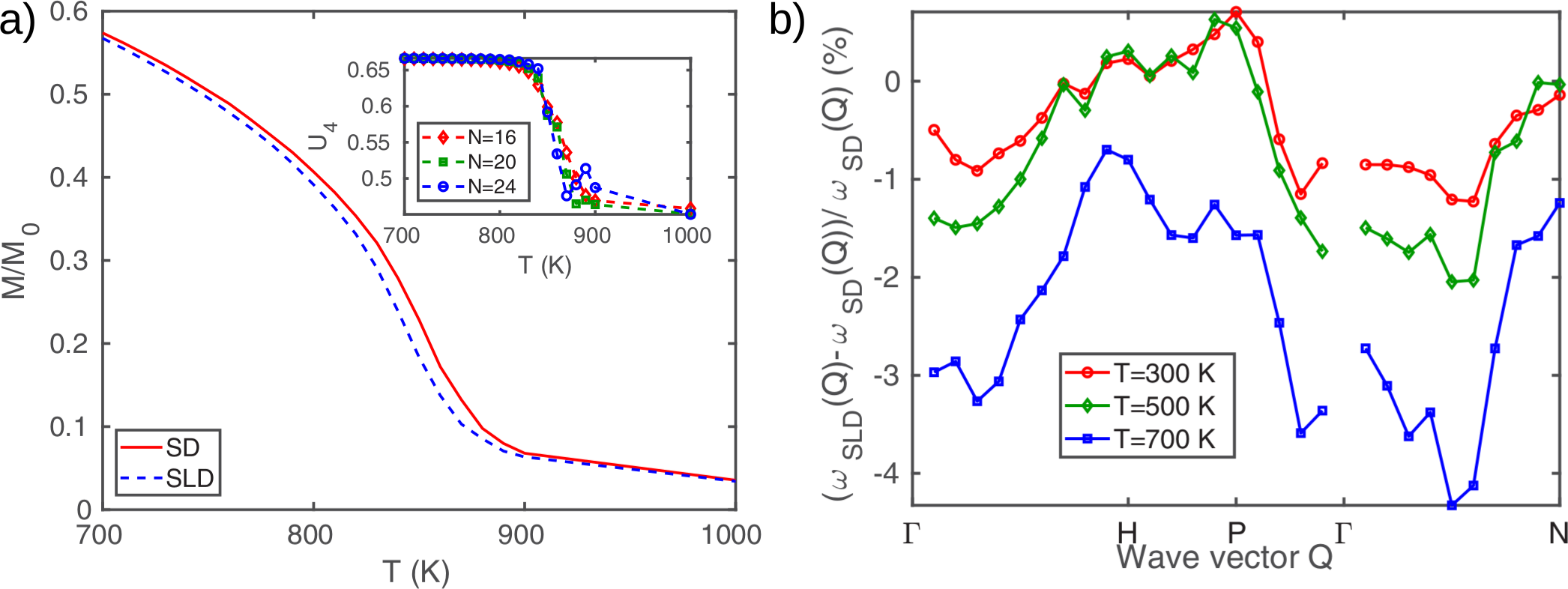}
}
\caption{a) The reduced magnetization of Fe \textit{bcc} at different temperatures modelled with pure spin (solid line) and spin-lattice (dashed line) dynamics; the inset show the scaling of the Binder constant with the system size, which is used for more accurate estimate of the ordering temperature ($T_\mathrm{C}=\unit[850]{K}$). b) The relative softening of the magnon frequencies due to spin-lattice interaction at different temperature; magnons are considered along a high-symmetry path in the Brillouin zone. Reproduced from \cite{Hellsvik2019}.}
\label{f:spin_lattice_Fe_bcc}
\end{figure}

Getting back to the more recent work \cite{Hellsvik2019}, it is interesting to see how combined spin-lattice atomistic simulations, using all the aforementioned equations (\ref{e:spin_Hamiltonian_HSS})--(\ref{e:dissipation_spin_and_lattice}) work in practice. The general methodology developed in that work is demonstrated, for example, for elemental ferromagnet Fe \textit{bcc}. Specific form of the spin-lattice coupling $-\frac{1}{2}\sum\limits_{ijk} \vec{A}_{ijk}\cdot \vec{u}_k (\vec{m}_i\cdot \vec{m}_j)$ corresponding to magnetostriction mechanism is considered in that work and all the Hamiltonian parameters are obtained fully from first principles using simulation cells with several thousands of atoms. From these calculations, an important conclusion was that one has to take into account both two-site ($i$, $j=k$) and three-site ($i$, $j\neq k$) terms and that the magnitude of the coupling coefficients $\vec{A}_{ijk}$ decays with distance slower than the Heisenberg exchange parameters $J_{ij}$. The coupled spin and lattice dynamics with these first-principles parameters showed, for example, that the magnetization of Fe is somewhat reduced due to spin-lattice effects, especially near the magnetic transition temperature (Fig.~\ref{f:spin_lattice_Fe_bcc}a). Magnon frequencies, which are calculated from the spin-spin dynamic structure factor, also show interesting effects due to spin-lattice coupling, such as the softening by a few percent for temperatures in the range $\unit[300-700]{K}$. It should be noted that the effect of thermal fluctuations that reduce the average magnetization lead to much larger magnon softening already in the pure spin dynamics simulations (see Fig.~17a in \cite{Hellsvik2019}).

The magnitude of spin-lattice effects in Fe \textit{bcc} is small but it can be more pronounced in systems close to structural or magnetostructural phase transition where the magnetic interactions and possibly even the macroscopic magnetic order are sensitive to structural details. As mentioned in \cite{Hellsvik2019}, possible candidate systems could be Laves compound YCo$_2$, Invar Fe-Ni alloy, and \textit{fcc} phase of Fe, where the later is known to host a variety of magnetic phases. Recently, another class of materials has been suggested as candidates for strong spin-lattice coupling, namely, the quasi-two-dimensional VOCl \cite{Komarek2009,Glawion2009,Ekholm2019} and similar compounds with Sc, Ti, Cr and Fe instead of V, and Br instead of Cl \cite{Zhang2008,Zhang2019,Qing2020,Zeng2022,Das2023}. Theoretical calculations in \cite{Amirabbasi2023} showed that this interesting physics is preserved also in the single-layered version of these systems.

\begin{figure}
{\centering
\includegraphics[width=0.99\textwidth]{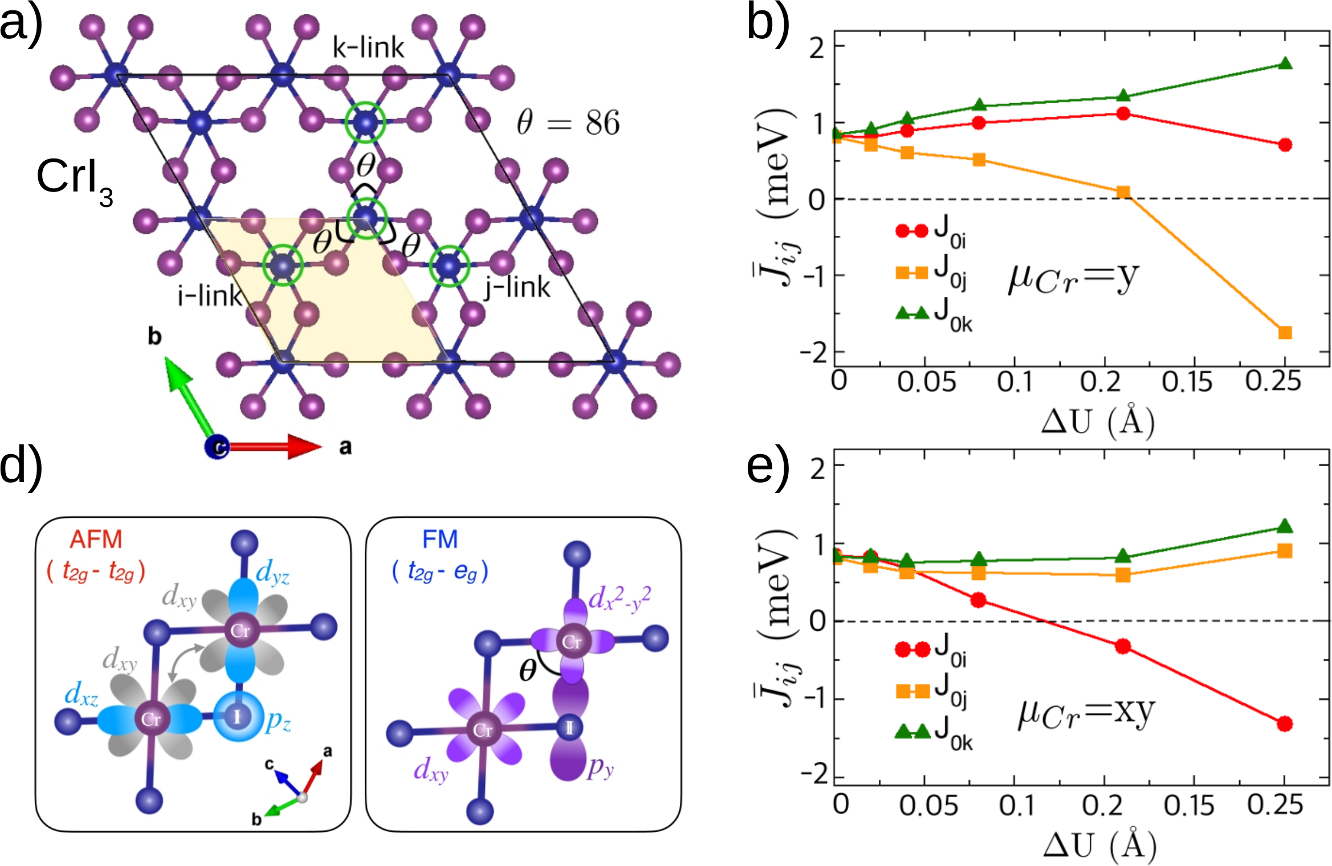}
}
\caption{a) The crystal structure of CrI$_3$ monolayer. b,e) The variation of Heisenberg interactions as functions of Cr displacement along different directions. d) Different orbital contributions to the Heisenberg interaction and the direct exchange between two Cr moments (grey orbitals). Figures are reproduced from \cite{Sadhukhan2022}.}
\label{f:CrI3}
\end{figure}
Another candidate for large spin-lattice coupling is CrI$_3$, which is one of the first discovered two-dimensional magnets and has topological magnon states \cite{Chen2018}. The later are due to the magnon gap $\sim\!\unit[4]{meV}$, also observed in theoretical simulations \cite{Kvashnin2020} but underestimated in magnitude. Spin-lattice coupling in this system has been discussed in terms of phonon spectra being sensitive to the magnetic structure \cite{Webster2018} as well as magnetic interactions changing in response to structural distortions \cite{Sadhukhan2022}. It appears that the frequencies of some of the Raman-active phonon modes in CrI$_3$ are different for the ferro- and antiferromagnetic configurations (Fig.~5ab in \cite{Webster2018}). This effect can be attributed to the difference of the electronic bands and character of the band gap for the two examples of possible magnetic configurations, as indicated by density functional theory calculations accompanying the experimental findings in \cite{Webster2018}. In another work \cite{Sadhukhan2022}, it was shown that Heisenberg interaction can even change the sign from FM to AFM for sufficiently large Cr displacements $\sim\!\unit[0.2]{\AA}$ (Fig.~\ref{f:CrI3}b,e). Similar effect under lattice strain is predicted in \cite{Luo2023}. Although average thermal displacements in the real system are not expected to reach this value (see Fig.~8 in \cite{Sadhukhan2022}), the magnitude of magnetic interactions will still change considerably during the coupled spin-lattice dynamics already at room temperature. The high complexity of CrI$_3$ magnet is also due to the sensitivity of Heisenberg and Dzyaloshinskii-Moriya interactions to displacements of non-magnetic ligands, not just the Cr sites. These spin-lattice effects can be understood in terms of the competition between AFM coupling of $t_{2g}$ orbitals and FM coupling of $t_{2g}$ and $e_{2g}$ orbitals \cite{Sadhukhan2022} (Fig.~\ref{f:CrI3}d).

\section{Conclusions}

We have discussed in this review how magnetic systems can be modelled on different length scales using the multiscale approach, starting from the electronic properties on the scale of individual atoms and electron hopping between them, proceeding with effective atomistic spin models representing magnetic excitations and textures in the system for a scale of up to several hundred nanometers and finishing with micromagnetic description of magnets in the continuous approximation where system sizes of several micrometers can be modelled. The important point is that all these different levels of multiscale modelling are connected to each other, because larger-scale models are derived from smaller-scale models, meaning that the model parameters are obtained from first principles and inherit the electronic structure feature of a given material. This improves the predictive power of the multiscale approach, as we have discussed on several examples from literature where the studied systems show interesting magnetic textures (spin spirals, skyrmions etc.). Furthermore, recent advances in modelling coupled spin-lattice phenomena were highlighted as well. The current successes and further development of the multiscale approach will ensure that many more magnetic systems and unusual phenomena will be discovered theoretically, leading to new applications and fundamental knowledge.

\section{Acknowledgements}
This work was financially supported by the Knut and Alice Wallenberg Foundation through grant numbers 2018.0060, 2021.0246, and 2022.0108 (Principal investigator: Prof. Dr. Olle Eriksson), and G\"oran Gustafsson Foundation (recipient of the ``small prize'': Vladislav Borisov). The author thanks Olle Eriksson, Anna Delin, Yaroslav O.~Kvashnin, Johan Hellsvik, Filipp N.~Rybakov, Ivan Miranda and Anders Bergman for very insightful discussions. The computations/data handling were enabled by resources provided by the Swedish National Infrastructure for Computing (SNIC) at the National Supercomputing Centre (NSC, Tetralith cluster) partially funded by the Swedish Research Council through grant agreement no.\,2018-05973 and by the National Academic Infrastructure for Supercomputing in Sweden (NAISS) at the National Supercomputing Centre (NSC, Tetralith cluster) partially funded by the Swedish Research Council through grant agreement no.\,2022-06725. Structural sketches in Figs.~\ref{f:multiscale_approach} and \ref{f:types_of_DMI} have been produced by the \textsc{VESTA3} software \cite{vesta2011}, while the spin configuration pictures in Figs.~\ref{f:types_of_DMI} were produced by the \textsc{Paraview} software \cite{paraview}.

\vspace{10pt}
\begin{center}
    {\large \textbf{References}}
\end{center}

\bibliographystyle{prb-titles}
\bibliography{main}

\end{document}